\documentclass[prd,preprintnumbers,superscriptaddress,nofootinbib,twocolumn]{revtex4-2}
\pdfoutput=1
\usepackage{enumerate}
\usepackage{amsmath,amssymb}
\usepackage{graphicx}
\usepackage{slashed}
\usepackage{cancel}
\usepackage{xspace,slashed}
\usepackage{float}
\usepackage[colorlinks,citecolor=blue, linkcolor = blue, urlcolor = blue]{hyperref}
\usepackage{titlesec}
\usepackage[normalem]{ulem}
\usepackage{subfigure,orcidlink}
\usepackage{multirow,array}
\usepackage{tabularray}
\UseTblrLibrary{booktabs}
\usepackage{bbding}% crossmark
\usepackage{charter}

\begin{document}

\title{Effective theory of light Dirac neutrino portal dark matter with observable ${\Delta N_{\rm eff}}$}

	%%%%%%%%%   Authors   %%%%%%%%%%%%

\author{{Debasish Borah\orcidlink{https://orcid.org/0000-0001-8375-282X}}}
\email{dborah@iitg.ac.in}
\affiliation{Department of Physics, Indian Institute of Technology Guwahati, Assam 781039, India}
\affiliation{Pittsburgh Particle Physics, Astrophysics, and Cosmology Center, Department of Physics and Astronomy, University of Pittsburgh, Pittsburgh, Pennsylvania 15260, USA}
\author{{Satyabrata Mahapatra\orcidlink{https://orcid.org/0000-0002-4000-5071}}}
\email{satyabrata@g.skku.edu}
\affiliation{Department of Physics and Institute for Basic Science, Sungkyunkwan University, Suwon 16419, South Korea}
\author{{Dibyendu Nanda\orcidlink{https://orcid.org/0000-0002-7768-7029}}}
\email{dnanda@het.phys.sci.osaka-u.ac.jp}
\affiliation{Department of Physics, Osaka University, Toyonaka, Osaka 560-0043, Japan}
    
\author{Sujit Kumar Sahoo{\orcidlink{https://orcid.org/0000-0002-9014-933X}}}
\email{ph21resch11008@iith.ac.in}
\affiliation{Department of Physics, Indian Institute of Technology Hyderabad, Kandi, Sangareddy 502285, Telangana, India}
    
\author{Narendra Sahu{\orcidlink{https://orcid.org/0000-0002-9675-0484}}}
\email{nsahu@phy.iith.ac.in}
\affiliation{Department of Physics, Indian Institute of Technology Hyderabad, Kandi, Sangareddy 502285, Telangana, India}

\begin{abstract}
We study the possibility of light Dirac neutrino portal dark matter (DM) in an effective field theory setup. Dirac nature of light neutrino automatically includes its right chiral part $\nu_R$ which, in our setup, also acts like a portal between DM and the standard model (SM) particles. Considering a Dirac fermion singlet DM stabilized by an unbroken $Z_2$ symmetry, we write down all possible dimension-6 effective operators involving DM-$\nu_R$ as well as $\nu_R$-SM which conserve $Z_2$, global lepton number and SM gauge symmetries. DM thermalization also ensures the thermalization of $\nu_R$, leading to enhanced effective relativistic degrees of freedom $N_{\rm eff}$, within reach of future cosmic microwave background (CMB) experiments. We study the complementarity among DM and CMB related observations for different Lorentz structures of effective operators. We also propose a UV complete gauged $\rm B-L$ symmetric model with Dirac neutrino portal dark matter.
\end{abstract}	
\maketitle
	%\flushbottom
	
\noindent

\section{Introduction}

The nature of dark matter (DM) and the origin of neutrino masses remain two of the most pressing questions in particle physics and cosmology \cite{ParticleDataGroup:2024cfk}. 
The existence of dark matter is strongly supported by cosmic microwave background (CMB) data and numerous astrophysical observations, including galaxy rotation curves, weak gravitational lensing, and the large-scale structure of the Universe. Recent data from the PLANCK satellite mission \cite{Planck:2018vyg} indicates that DM constitutes about 26.8\% of the total energy density of the Universe. However, none of the standard model (SM) particles meets all the necessary criteria to be a viable DM candidate, leading to the development of numerous beyond the Standard Model (BSM) frameworks.

Similarly, neutrino oscillation experiments~\cite{Super-Kamiokande:1998kpq, SNO:2001kpb, DoubleChooz:2011ymz, DayaBay:2012fng, RENO:2012mkc} have revealed that neutrinos possess tiny masses and exhibit large leptonic mixing, which can not be accounted for in the SM where neutrinos remain massless. While these experiments have provided crucial insights regarding the mixing angles and mass squared differences, they remain insensitive to the fundamental nature of neutrinos, whether they are Dirac or Majorana fermions. Although the experiments searching for neutrinoless double-beta decay could potentially confirm the Majorana nature of neutrinos, no such observations have been made to date~\cite{ParticleDataGroup:2024cfk}. This has sparked growing interest in exploring the possibility that light neutrinos could be Dirac fermions, challenging the long-held focus on Majorana neutrinos in conventional neutrino mass models~\cite{Minkowski:1977sc, Gell-Mann:1979vob, Mohapatra:1979ia,Sawada:1979dis,Yanagida:1980xy, Schechter:1980gr, Mohapatra:1980yp, Schechter:1981cv, Wetterich:1981bx, Lazarides:1980nt, Brahmachari:1997cq, Foot:1988aq, Dutta:2020xwn, Borah:2021rbx, Borah:2022zim, Konar:2020wvl, Konar:2020vuu, Bhattacharya:2017sml,Bhattacharya:2016rqj,Borah:2024wos,Paul:2024iie,Coito:2022kif}.

In light of these open questions, there is growing interest in models that connect dark matter and neutrino physics, potentially providing a unified solution to these fundamental puzzles~\cite{Yao:2017vtm,Carvajal:2018ohk,Nanda:2019nqy, Borah:2020jzi,Das:2023yhv}. One such approach involves the concept of neutrino portal dark matter, where either SM light neutrinos or heavy neutrinos serve as a mediator between the dark sector and the visible sector~\cite{Falkowski:2009yz, GonzalezMacias:2015rxl, Batell:2017rol, Batell:2017cmf, Bandyopadhyay:2018qcv, Chianese:2018dsz, Blennow:2019fhy, Lamprea:2019qet, Chianese:2019epo, Bandyopadhyay:2020qpn, Hall:2019rld, Berlin:2018ztp, Xu:2023xva,Ahmed:2023vdb}. Due to the absence of tree level DM-nucleon scattering, this framework is particularly intriguing in light of the null results from direct DM searches. Direct search experiments such as PandaX-II~\cite{PandaX-II:2017hlx}, XENON-nT~\cite{XENON:2023cxc,XENON:2019zpr}, and LUX-ZEPLIN~\cite{LZ:2022lsv} have probed spin-independent DM-nucleon cross sections down to ~$10^{-48}~{\rm GeV}^{-2}$ for DM masses ranging from a few GeV to $\mathcal{O}(10^4)$ GeV with weaker constraints for lighter DM masses from experiments like DarkSide~\cite{DarkSide:2018bpj}, CDMSlite~\cite{SuperCDMS:2018gro}, and CRESST-III~\cite{CRESST:2019jnq}. Similar bounds also exist for DM-electron scattering \cite{XENON:2019gfn}. These null detections motivate the exploration of novel scenarios like neutrino portal DM and their search strategies.

Motivated by these, here we consider right chiral neutrinos $(\nu_R)$ introduced to explain light Dirac neutrino mass act like a portal between DM and the visible sector. In earlier works \cite{Biswas:2021kio, Biswas:2022vkq, Das:2023yhv}, light Dirac neutrino portal DM was studied by considering specific models. In the present work, we adopt a more general effective field theory (EFT) approach, focusing on dimension-6 operators that respect relevant symmetries like the SM gauge symmetries, accidental global symmetries like lepton and baryon numbers and a $Z_2$ symmetry protecting the stability of DM, chosen to be a vectorlike singlet fermion $\chi$. The effective interactions of such $\nu_R$-philic DM with $\nu_R$ and interactions of $\nu_R$ with SM arise at the order of dimension-6 or above. We consider such dimension-6 operators of all possible Lorentz structures. There exist no DM-SM operators at dimension-6 in the spirit of $\nu_R$-philic DM, keeping direct detection rates suppressed, in agreement with experimental results.

A key aspect of this scenario is the potential impact on cosmological observables, particularly the effective number of relativistic species ($N_{\rm eff}$). The thermalization of right-handed neutrinos in the early Universe can lead to an enhancement in $N_{\rm eff}$, potentially detectable by future CMB experiments. This connection between particle physics and cosmology offers a unique opportunity to probe the dark sector and neutrino properties simultaneously.
Our study aims to explore the complementarity between dark matter phenomenology and CMB observations, considering various Lorentz structures of the effective operators. We also propose a gauge extension of the SM: the $\rm B-L$ model. This UV completion provide concrete realizations of the EFT framework and offer additional avenues for experimental tests.

In the following sections, we present our EFT setup, analyze the resulting phenomenology, and discuss implications for future experiments and theoretical model-building in the context of neutrino portal dark matter. Section \ref{subsec:deltaN} discusses the temperature evolution of $\nu_R$ and its contribution to $\Delta N_{\rm eff}$, followed by an analysis of DM relic abundance and loop-suppressed direct detection signatures in Sec. \ref{subsec:dm}. Section \ref{sec:uvmodels} examines the parameter space of UV-complete gauged $B-L$ model considering $\Delta N_{\rm eff}$ bounds and finally we conclude in Sec. \ref{sec:conclusion}.

\section{Effective field theory approach to $\Delta N_{\rm eff}$ and dark matter}\label{sec:EFT}

The EFT approach serves as a powerful tool to study interactions involving new particles without committing to a specific ultraviolet (UV) completion. By parametrizing the effects of heavy mediators through higher-dimensional operators, EFT provides a model-independent framework that captures a wide range of phenomenological possibilities. This is particularly useful for scenarios where the new physics scale is significantly higher than the energies accessible in current experiments or where direct detection is not feasible. The dark matter phenomenology in such an EFT setup can be discussed with very few free parameters, namely, DM mass, cutoff scale, and the corresponding coupling or Wilson coefficients. Such DM EFT has been studied extensively in the context of direct detection, indirect detection as well as collider searches in several works \cite{Beltran:2008xg, Fan:2010gt, Goodman:2010ku, Beltran:2010ww, Fitzpatrick:2012ix}, also summarized in a recent review \cite{Bhattacharya:2021edh}.

In this study, we extend the SM particle content by introducing three copies of right-handed neutrinos ($\nu_R$) constituting Dirac neutrinos with $\nu_L$ in SM and a massive vectorlike fermion ($\chi$) that is a candidate for DM with mass $M_{\rm DM}(\equiv M_\chi)$. Both $\nu_R$ and $\chi$ are singlets under the SM gauge group. The vectorlike fermion $\chi$ is stabilized by an additional unbroken $Z_2$ symmetry, under which all SM particles, including $\nu_R$ are even. In this framework, $\nu_R$
carries a leptonic charge similar to SM leptons, keeping light neutrinos purely Dirac in the limit of unbroken global lepton number symmetry. We consider a subclass of leptophilic DM EFT with light Dirac neutrinos \cite{Borah:2024twm} where DM is only ‘$\nu_R$-philic' {\it i.e.} $\chi$ interacts exclusively with $\nu_R$ via EFT operators, making $\nu_R$ a portal between the visible sector and the dark sector. While this keeps direct detection cross section of DM highly suppressed at one-loop, it opens up other detection aspects as we discuss in the upcoming sections.
We emphasize that our assumption of vanishing interactions among $\chi$ and SM leptons $f$ requires additional symmetries in a UV complete setup. Such symmetries can either forbid $\chi-f$ interactions at all orders or at the leading order. In the latter scenario, such operators involving $\chi-f$ interactions are generated only at loop level and hence are suppressed in our EFT analysis. Such a possibility can indeed be achieved in UV complete scenario which we discuss in Sec. \ref{sec:uvmodels}.

In a minimal setup, we propose the following EFT Lagrangian:

\begin{eqnarray}\label{eq:effLag2}
-\mathcal{L}&\supset& {G}_{S}\overline{f_{L}}\nu_{R}\overline{\nu_{R}}f_{L}+G_{V} \overline{f_{L}}\gamma^{\mu}f_{L}\overline{\nu_{R}}\gamma_{\mu}\nu_{R}
    \nonumber\\&
    {}&+G_T(\overline{L} \sigma^{\mu\nu} \nu_R \tilde{H} B_{\mu\nu}+\overline{L} \sigma^{\mu\nu} \nu_R \tilde{H}\sigma_i W^i_{\mu\nu})\nonumber\\&{}&+G'_V\overline{\chi_{x}}\gamma^{\mu}\chi_{x}\overline{\nu_{R}}\gamma_{\mu}\nu_{R}+{G'_S}\overline{\chi_{L}}\nu_{R}\overline{\nu_{R}}\chi_{L}
\end{eqnarray}

Using Fierz transoformation, this can be written as
\begin{eqnarray}\label{eq:effLag2FR}
-\mathcal{L}&\supset& (G_{V}- k~G_{S}) \overline{f_{L}}\gamma^{\mu}f_{L}\overline{\nu_{R}}\gamma_{\mu}\nu_{R}
    \nonumber\\&
    {}&+G_T(\overline{L} \sigma^{\mu\nu} \nu_R \tilde{H} B_{\mu\nu}+\overline{L} \sigma^{\mu\nu} \nu_R \tilde{H}\sigma_i W^i_{\mu\nu})\nonumber\\&{}&+(G'_V- k' G'_{S})\overline{\chi_{L}}\gamma^{\mu}\chi_{L}\overline{\nu_{R}}\gamma_{\mu}\nu_{R}\nonumber\\&{}&+{G'_V}\overline{\chi_{R}}\gamma^{\mu}\chi_{R}\overline{\nu_{R}}\gamma_\mu\nu_{R}
\end{eqnarray}

where $x\in{L,R}$ and
$G_S,G_V,G_T,G'_S, G'_V$ are the dimensionful effective coefficients of the dimension-6 operators. The dimensionless parameters, $k$ and $k'$ carry a value of 1/2 \cite{Aebischer:2022aze}. The first three terms describe $\nu_{R}$-SM interactions that can give rise to number changing and elastic scattering processes of $\nu_R$ with SM fermions $f_{L,R}$ and lepton doublet, $L$, whereas the last two terms represent the interaction between DM and $\nu_R$. Since DM interacts only with $\nu_R$ at leading order while $\nu_R$ interacts with both SM and DM at the same order, it justifies the $\nu_R$-philic nature of DM. Furthermore, the tensor type $\nu_R$-SM effective interaction, with coefficient $G_T$, can give rise to neutrino magnetic moment (NMM). The measured upper bound on the NMM ($\mu_\nu$) from electron scattering experiment is approximately $\mu_\nu\sim 2.9\times10^{-11}\mu_B$ \cite{Beda:2012zz}, where $\mu_B=\frac{e}{2m_e}$ is the Bohr magneton. More recent data from laboratory experiments such as XENONnT \cite{XENON:2022ltv} and Borexino \cite{Borexino:2017fbd} have provided a significantly stricter limit on the NMM, yielding $\mu_\nu\sim 6.4\times10^{-12}\mu_B$ and $\mu_\nu\sim 28\times10^{-12}\mu_B$, respectively. In contrast, the astrophysical bound derived from the red giant branch \cite{Capozzi:2020cbu}, place a much stronger constraint, $\mu_\nu \sim 1.5 \times 10^{-12} \mu_B$.

\subsection{EFT approach to $\Delta N_{\rm eff}$}\label{subsec:deltaN}
The thermalization of $\nu_R$ contributes to the effective relativistic degrees of freedom defined as
\begin{eqnarray}
 N_{\rm eff} \equiv \frac{8}{7} \left( \frac{11}{4} \right)^{4/3} \left( \frac{\rho_{\rm Tot} -\rho_{\gamma}}{\rho_{\gamma}} \right),     
\end{eqnarray}
where $\rho_{\rm Tot}$ is the total radiation content of the Universe with $\rho_\gamma$ being the photon energy density. CMB measurements constrain such additional relativistic degrees of freedom as ${\rm N_{eff}= 2.99^{+0.34}_{-0.33}}$ at $2\sigma$ or $95\%$ CL including baryon acoustic oscillation (BAO) data \cite{Planck:2018vyg}. The translated bound on $\Delta N_{\rm eff}(={\rm N}_{\rm eff}-{\rm N}^{\rm SM}_{\rm eff})$ at $2\sigma$ can be written as $\Delta N_{\rm eff} \lesssim 0.285$, where $N_{\rm eff}^{\rm SM}=3.045$\cite{Mangano:2005cc, Grohs:2015tfy,deSalas:2016ztq}. The latest DESI 2024 data give a slightly weaker bound $\Delta N_{\rm eff} \lesssim 0.4$ at $2\sigma$ CL \cite{DESI:2024mwx}. The recent result from ACT \cite{ACT:2025tim} combined with PLANCK and the measurements of primordial deuterium and helium is also consistent with a positive contribution to $\Delta N_{\rm eff}<0.075$ at $2\sigma$ CL. Similar bound also exists from big bang nucleosynthesis (BBN) $2.5 < {N}_{\rm eff} <3.2$ at $95\%$ CL \cite{Fields:2019pfx}. These cosmological bounds are consistent with the SM predictions. Future CMB experiment CMB-S4 is expected to reach a much better sensitivity of $\Delta {\rm N}_{\rm eff}
= 0.06$ \cite{Abazajian:2019eic}, taking it closer to the SM prediction. Another future experiment CMB-HD \cite{CMB-HD:2022bsz} can probe $\Delta N_{\rm eff}$ up to $0.027$ at $2\sigma$.

There have been several recent works on enhanced $N_{\rm eff}$ with Dirac neutrinos \cite{Abazajian:2019oqj, FileviezPerez:2019cyn, Nanda:2019nqy, Han:2020oet, Luo:2020sho, Borah:2020boy, Adshead:2020ekg, Luo:2020fdt, Mahanta:2021plx, Du:2021idh, Biswas:2021kio, Borah:2022obi, Borah:2022qln, Li:2022yna, Biswas:2022fga, Adshead:2022ovo, Borah:2023dhk, Borah:2022enh, Das:2023oph, Esseili:2023ldf, Das:2023yhv, Angel:2025luo}. EFT of light Dirac neutrino interactions has also been studied in the context of $N_{\rm eff}$ \cite{Luo:2020sho, Luo:2020fdt}. In another recent work \cite{Borah:2024twm}, EFT of leptophilic DM with light Dirac neutrinos was studied in the context of collider and $N_{\rm eff}$ phenomenology. In another recent work \cite{Biswas:2024gtr}, EFT of Dirac neutrino phenomenology was studied up to dimension-6. Here, we consider a similar EFT setup where DM is $\nu_R$-philic uniting thermalization of DM and $\nu_R$ with the visible sector.

To study the evolution of $\nu_R$ energy density, we consider the continuity equation for a homogeneous and isotropic universe:
\begin{equation}
    \dot{\rho}_{\rm Tot}+3\mathcal{H}(\rho_{\rm Tot}+P_{\rm Tot})=0,
\end{equation}
where $\mathcal{H}=\sqrt{\frac{8\pi}{3M^{2}_{Pl}}\rho_{\rm Tot}}$ is the Hubble parameter, and $\rho_{\rm Tot}$ and $P_{\rm Tot}$ are the total energy density and pressure, respectively.  These can be decomposed into contributions from the SM bath and $\nu_R-\chi$ bath as
\begin{eqnarray}
    \rho_{\rm Tot}&=&\rho_{\rm SM}+(\rho_{\nu_R}+\rho_{\chi}),\\
    P_{\rm Tot}&=&P_{\rm SM}+(P_{\nu_R}+P_{\chi}).
\end{eqnarray}
The evolution of two separate sectors namely, $\nu_R$-DM and the SM bath is described by a coupled system as $\nu_R$ couples to the SM via similar dimension-6 operators. The effective coupling for $\chi-\nu_R$ interaction is chosen to be larger than $0.1\times G_F$, ensuring that the DM remains in thermal equilibrium with the $\nu_R$ bath in the early Universe.
While these baths can maintain thermal equilibrium with each other, observational constraints from PLANCK 2018 indicate that for three $\nu_R$ species, they must have decoupled from the SM plasma at temperatures above $600$ MeV \cite{Abazajian:2019oqj} -- much earlier than the decoupling of left-handed neutrinos ($\nu_L$). When the $\nu_R$-SM interaction rate drops below the Hubble rate, the two baths decouple and evolve independently according to
\begin{eqnarray}
\dot{\rho}_{\nu_R}+\dot{\rho}_{\chi}+3\mathcal{H}(\rho_{\nu_R}+P_{\nu_R})+3\mathcal{H}(\rho_{\chi}+P_{\chi})=C^{(\rho)}_{\nu_R},\label{eq:evolnuR}\\
    \dot{\rho}_{\rm SM}+3\mathcal{H}(\rho_{\rm SM}+P_{\rm SM})=-C^{(\rho)}_{\nu_R}.\label{eq:evolSM}
\end{eqnarray}
Here, $C^{(\rho)}_{\nu_R}$ represents the collision term that quantifies energy transfer between the baths through annihilation and scattering processes. For a generic process $1+2\leftrightarrow3+4$, this term is given by
{\scriptsize{
\begin{eqnarray}
    C^{(\rho)}_{\nu_R}=&-&N_{\nu_R}\int E_1 d\Pi_1d\Pi_2d\Pi_3d\Pi_4(2\pi)^4\delta^4(p_1+p_2-p_3-p_4) S\nonumber\\
    &\times&\left[|\mathcal{M}|^2_{1+2\rightarrow3+4}f_1f_2(1-f_3)(1-f_4)\right.\nonumber\\
    &{}&\left.-|\mathcal{M}|^2_{3+4\rightarrow1+2}f_3f_4(1-f_1)(1-f_2)\right],
\end{eqnarray}
}}
where
\begin{equation}
    d\Pi_i=\frac{g_i}{(2\pi)^3}\frac{d^3p_i}{2E_i},~~~f_i=\frac{1}{e^{E_i/T_i}+1},~~(i=1,2,3,4).
\end{equation}
In these equations, $N_{\nu_R}$ represents the degrees of freedom for $\nu_R$ species, and $S$ is the symmetry factor for the relevant processes. Each particle $i$ is characterized by its internal degrees of freedom ($g_i$), energy ($E_i$), and temperature ($T_i$). Here, it is worth mentioning that, for temperatures around and above the $\nu_R$-SM decoupling temperature ($\gtrsim$ 500 MeV), the collision term is dominated by interactions with light SM fermions (electrons and neutrinos), whose masses can be safely neglected at these energy scales.

The evolution equations namely, Eq. (\ref{eq:evolnuR}) and Eq. (\ref{eq:evolSM}), can be reformulated to show the temperature evolution of $\nu_R$ relative to the SM bath temperature as
\begin{eqnarray}\label{eq:tnurevo}
	\frac{d(\rho_{\nu_R}+\rho_\chi)}{d\rho_{\rm SM}}&=&\frac{3\mathcal{H}(\rho_{\nu_R}+P_{\nu_R})+3\mathcal{H}(\rho_{\chi}+P_{\chi})-C^{(\rho)}_{\nu_R}}{3\mathcal{H}(\rho_{\rm SM}+P_{\rm SM})-C^{(\rho)}_{\nu_R}},\nonumber\\
	\frac{dT_{\nu_R}}{dT_{\gamma}}&=&\frac{3\mathcal{H}(\rho_{\nu_R}+P_{\nu_R})+3\mathcal{H}(\rho_{\chi}+P_{\chi})-C^{(\rho)}_{\nu_R}}{3\mathcal{H}(\rho_{SM}+P_{SM})-C^{(\rho)}_{\nu_R}}\nonumber\\
	&{}&\times\frac{\partial\rho_{SM}}{\partial T_\gamma}\left(\frac{\partial\rho_{\nu_R}}{\partial T_{\nu_R}}+\frac{\partial\rho_{\chi}}{\partial T_{\nu_R}}\right)^{-1}.
\end{eqnarray}

\begin{figure}[h]
		\centering
        \includegraphics[scale=0.45]{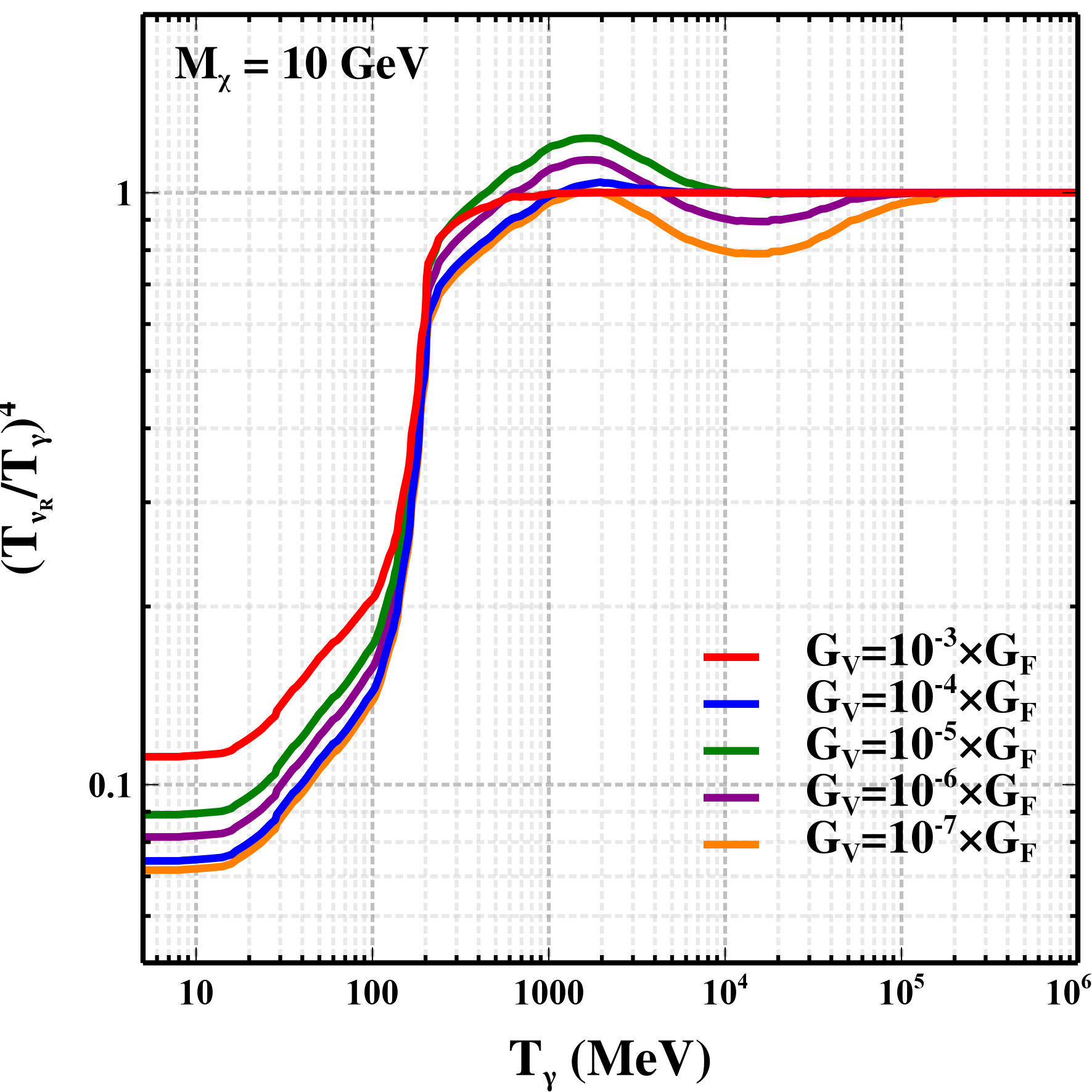}
		\caption{The evolution of $(T_{\nu_R}/T_\gamma)^4$ for several benchmark values of $G_V$ with $M_{\chi}=10$ GeV.}
		\label{fig:tempnuR}
\end{figure}

We have calculated the evolution of right-handed neutrino temperature ($T_{\nu_R}$) in the early Universe, focusing on the period when the SM bath temperature ($T_\gamma$) cooled from $10^6$ MeV down to $10$ MeV. The expressions for $C^{(\rho)}_{\nu_R}$ corresponding to the $\nu_R$-SM interactions are provided in Appendix \ref{app:EFTnuR}. We assume the right-handed neutrinos to be thermally equilibrated with the SM bath particles initially. Figure \ref{fig:tempnuR} shows the temperature ratio $(T_{\nu_R}/T_\gamma)^4$ versus $T_\gamma$ for different choices of the effective four-fermion interaction strength as mentioned in the inset of the same figure. We show the temperature evolution only for vector-like $\nu_R$-SM coupling $G_V$ and denote its strength relative to the Fermi coupling $G_F$, where we considered the DM mass to be 10 GeV. As previously mentioned, here the $\chi-\nu_R$ coupling $G'_{V}$ is chosen such that $\chi$ remains in thermal equilibrium with the $\nu_R$ bath. Deviations from the $(T_{\nu_R}/T_\gamma)^4=1$ at lower temperatures indicate the decoupling of $\nu_R$ bath from the SM bath. The plot shows that a smaller value of $G_V$ leads to an early decoupling of $\nu_R$ bath, as expected. In addition to the effective coefficients, the $T_{\nu_R}$ evolution also depends on DM mass. As noted previously, the DM interacts solely with the $\nu_R$ bath. Therefore, when $T_{\nu_R}$ falls below the DM mass, the DM transfers its entropy to the $\nu_R$ thermal bath. At this epoch, if the $\nu_R$ bath remains in equilibrium with the SM thermal bath, the ratio $T_{\nu_R}/T_\gamma$ will remain constant due to quick dissipation of $\nu_R$ energy into the SM bath. However, if the $\nu_R$ thermal bath has decoupled from the SM thermal bath, the entropy discharge from the DM leads to an increase in $\nu_R$ bath temperature. This behavior is clearly shown in Fig. \ref{fig:tempnuR}. The DM of mass 10 GeV transfers its entropy during the epoch of 1 GeV $\lesssim T_\gamma \lesssim$ 10 GeV. The purple and orange colored contours correspond to the decoupling of $\nu_R$ from the SM bath before $T_\gamma\simeq10^4$ MeV, and receive the maximum entropy contribution from the DM, leading to an increase in $T_{\nu_R}$ once $T_{\nu_R}\simeq M_{\chi}$. In contrast, the red colored contour shows $\nu_R$-SM decoupling at $T_{\nu_R}\simeq 1$ GeV and does not receive any further contribution from DM entropy transfer. The green and blue contours show $\nu_R$-SM decoupling in the same epoch when DM transfers entropy to $\nu_R$ bath. Between these two contours, the green contour corresponds to earlier $\nu_R$-SM decoupling, leading to a larger  $T_{\nu_R}/T_\gamma$ ratio, as it receives a greater contribution from the DM entropy transfer compared to the blue contour. We also find similar behavior of $T_{\nu_R}$ evolution for both scalar and tensor type interactions.

The contribution of $\nu_R$ to the effective number of neutrino species $N_{\rm eff}$ can be expressed as
{\scriptsize{
\begin{eqnarray}
    \Delta N_{\rm eff}=N_\nu\left(\frac{11}{4}\right)^{4/3}\frac{\rho_{\nu_R,0}}{\rho_{SM,0}}&=&N_\nu\left(\frac{11}{4}\right)^{4/3}\frac{T^4_{\nu_R,0}}{T^4_{\gamma,0}}\label{eq:deln0}\\
    &=&N_\nu\left(\frac{T_{\nu_R,10}}{T_{\gamma,10}}\right)^4\,,\label{eq:deln10}
\end{eqnarray}
}}
where subscript “0" refers to the CMB formation temperature and “10" indicates a photon bath temperature of 10 MeV. $N_\nu$ represents the number of $\nu_R$ species present in the model.

\begin{figure}[h]
		\centering
        \includegraphics[scale=0.45]{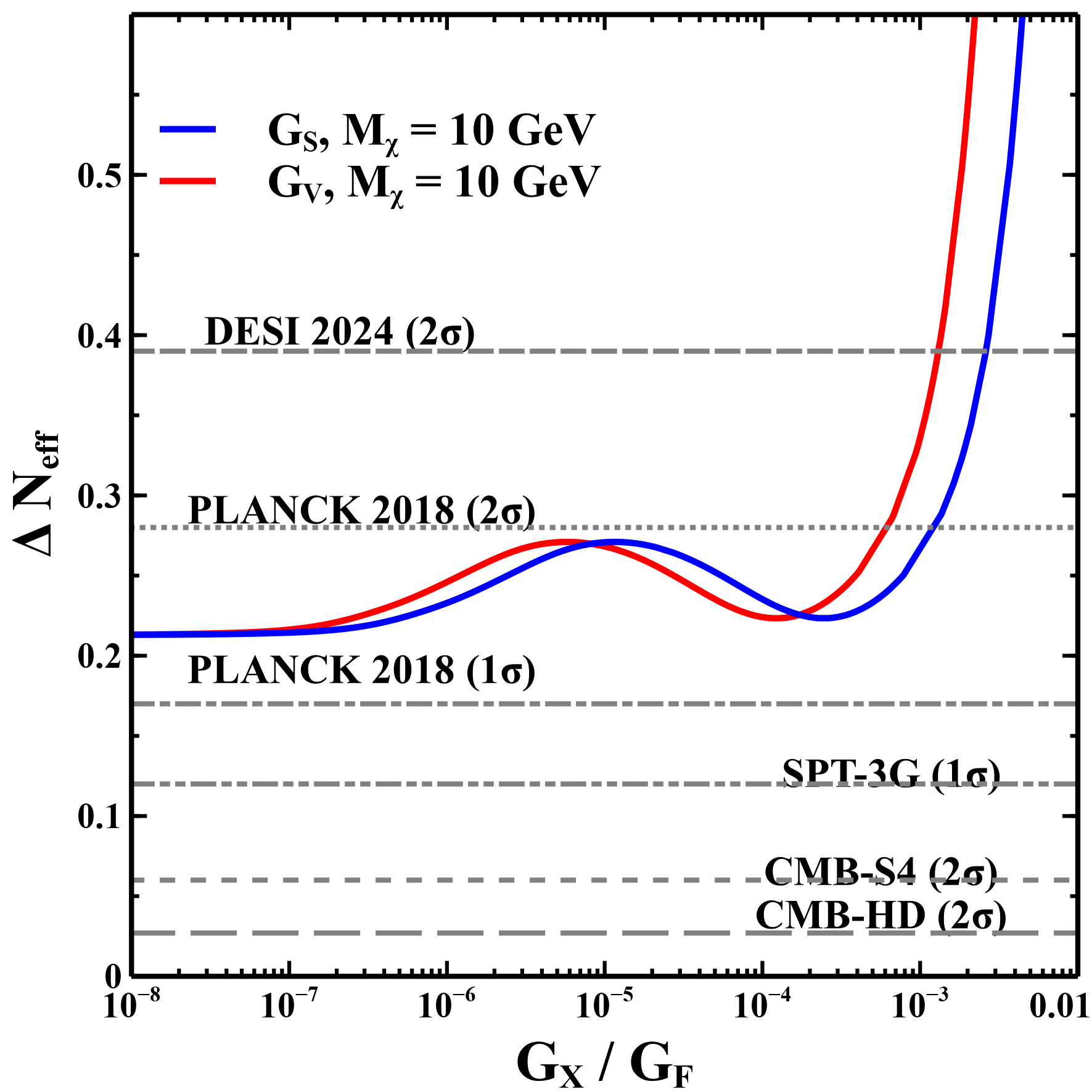}
		\caption{ $\Delta N_{\rm eff}$ as a function of $ {G_X \, (X\equiv S, V)}$ with $M_{\chi}=10$ GeV. We have shown the current upper bounds from PLANCK 2018 \cite{Planck:2018vyg} and the analysis by \cite{Allali:2024cji} using the recent DESI 2024 data. The sensitivity of upcoming experiments SPT-3G \cite{SPT-3G:2019sok}, CMB-S4 \cite{Abazajian:2019eic} and CMB-HD \cite{CMB-HD:2022bsz} are also shown.}
		\label{fig:neffSV}
\end{figure}
\begin{figure}[h]
		\centering
		\includegraphics[scale=0.45]{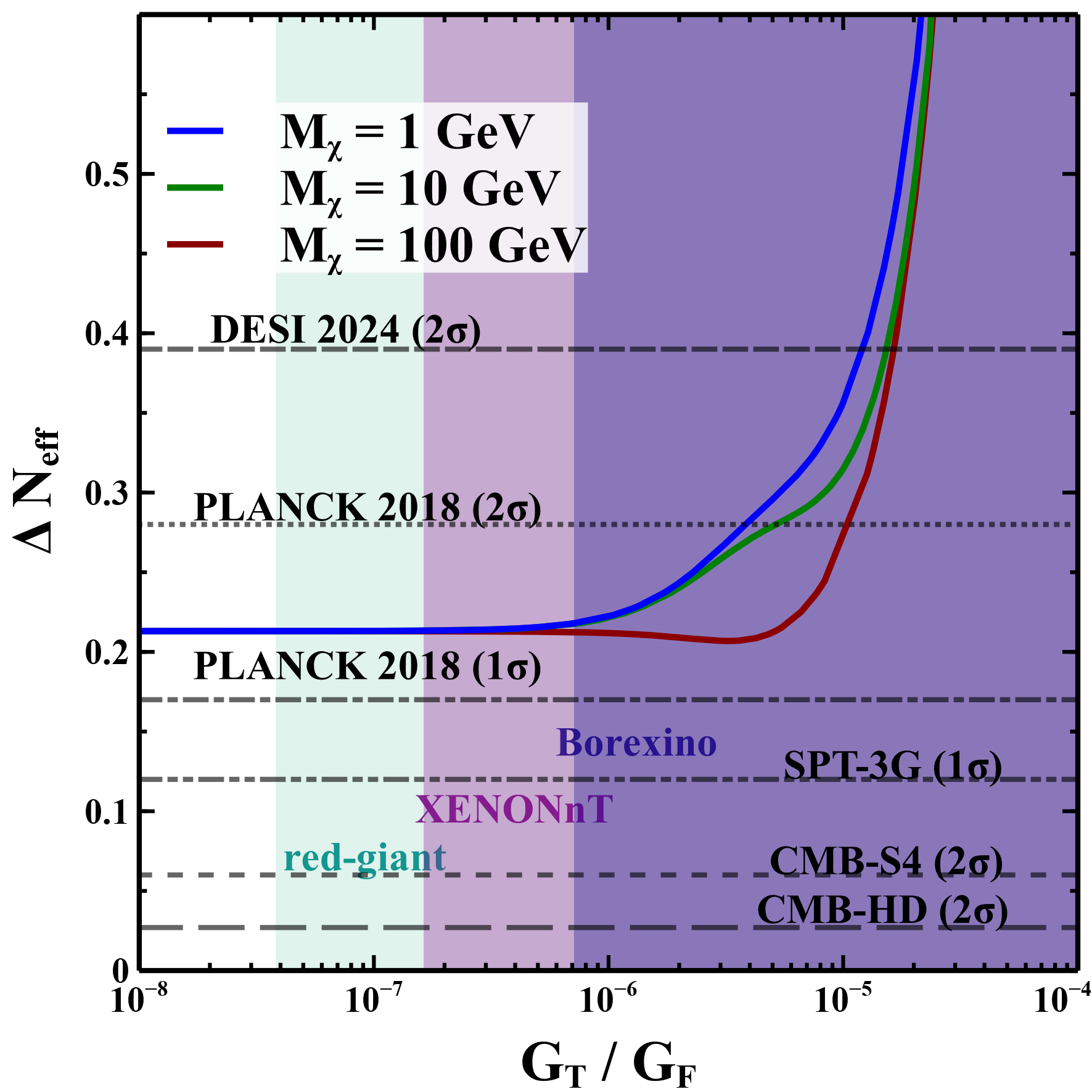}
		\caption{$\Delta N_{\rm eff}$ as a function of $G_T$ is shown for $M_{\chi} = (1, 10, 100)$ GeV. The shaded regions represent the values of $G_T$ that are excluded by the NMM constraints. The current upper bounds and future sensitivities from CMB experiments are indicated by the horizontal gray lines.}
		\label{fig:neffT}
\end{figure}
In Fig. \ref{fig:neffSV}, we showcase the $\Delta N_{\rm eff}$ due to $\nu_R$ as a function of the effective couplings $G_S$ and $G_V$, for $M_{\chi}=10$ GeV. In the horizontal axis, the strength of $G_S$ and $G_V$ are shown relative to the Fermi constant $G_F$. For $G_X<10^{-5}\times G_F$, $\nu_R$ decouples before $T_\gamma\simeq10$ GeV, resulting in an equal contribution to $\Delta N_{\rm eff}$ with a minimum value $\Delta N_{\rm eff} \approx 0.21$. It should be noted that the minimum value of $\Delta N_{\rm eff}$ may vary depending on the DM mass, in general. A dip is observed around $G_X\simeq 10^{-4}\times G_F$, which results from the interplay between the epochs of the $\nu_R$ decoupling and entropy transfer from DM. We have shown the existing constraints from PLANCK 2018 and recent results of DESI 2024. We also project the future sensitivities of SPT-3G, CMB-S4 and CMB-HD. While PLANCK 2018 results rule out $G_S/G_F$ and $G_V/G_F$ larger than $5.7\times10^{-6}$, DESI 2024 results allows $G_S$ and $G_V$ as large as $5.6\times10^{-4}$. Notably, the projected sensitivities of future experiments like SPT-3G, CMB-S4, and CMB-HD can completely probe these scenarios in which $\nu_R$ is thermalized with SM bath in the early Universe.
Similarly, Fig. \ref{fig:neffT} illustrates the contribution to $\Delta N_{\rm eff}$ arising from the $\nu_R$-SM tensorial interaction term as a function of the respective coupling $G_T$. The variation in $\Delta N_{\rm eff}$ for $M_{\chi} = (1, 10, 100)$ GeV is shown as colored contours. The horizontal gray lines represent the current bounds and future sensitivities from CMB experiments, while the shaded regions indicate the exclusion limit due to upper bound on Dirac NMM from experiments such as XENONnT \cite{XENON:2022ltv}, Borexino \cite{Borexino:2017fbd} and astrophysical observations related to red-giants \cite{Capozzi:2020cbu}. We have used these limits to constrain the effective tensorial operator, $G_T$, and it excludes $G_T\geq 3.82 \times 10^{-8} G_F$. Due to the presence of DM interacting via $\nu_R$-portal, we get enhanced contribution to $N_{\rm eff}$ compared to a scenario without DM \cite{Li:2022dkc}.

\subsection{Phenomenology of $\nu_R$-philic dark matter}\label{subsec:dm}
\begin{enumerate}
	\item \textit{DM Relic Density:}\\
In this section, we explore the relic density of vector-like fermionic dark matter ($\chi$), which interacts exclusively with right-handed neutrinos ($\nu_R$ ) through dimension-6 effective operators. These interactions are crucial in determining the annihilation of DM into $\nu_R$ as shown in Fig. \ref{fig:c2annihilation}
and the resulting relic density. The relevant terms in the EFT Lagrangian include a scalar-type interaction, governed by the effective coupling $G'_S$ ({\it i.e.,} ${G'_S}\overline{\chi_{L}}\nu_{R}\overline{\nu_{R}}\chi_{L}$) and
a vector-type interaction proportional to effective coupling $G'_V$ ({\it i.e.} $G'_V\overline{\chi_{x}}\gamma^{\mu}\chi_{x}\overline{\nu_{R}}\gamma_{\mu}\nu_{R}$ where $x \equiv L,R$ ). In a UV complete framework, the scalar interaction can contribute to a $t$-channel annihilation process of the DM, whereas the vector interaction can contribute to a $s$-channel annihilation of DM into $\nu_R$.  

\begin{figure}[h]
		\centering
  \includegraphics[scale=0.1]{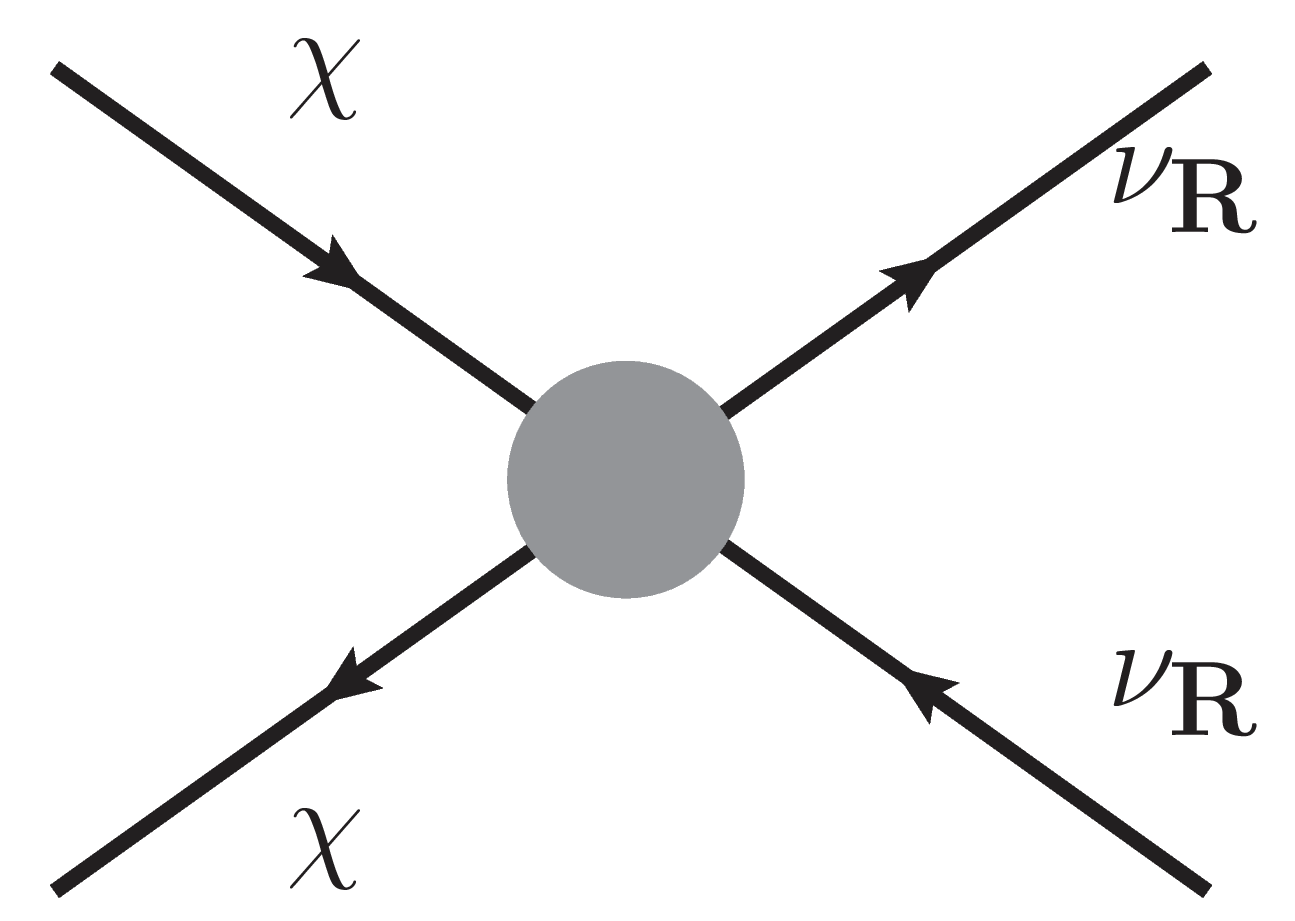}
		\caption{DM annihilation to $\nu_R$.}
		\label{fig:c2annihilation}
\end{figure}

In Sec. \ref{subsec:deltaN}, we discussed the evolution of the $\nu_R$ thermal bath temperature relative to the SM temperature, as illustrated in Fig. \ref{fig:tempnuR}. As DM $\chi$ exclusively interacts with $\nu_R$, the thermal evolution of $\nu_R$ bath directly impacts the dynamics of DM freeze-out. Notably, it can give rise to two distinct scenarios as follows:
\begin{enumerate}
\item Early freeze-out: If dark matter freezes out before $\nu_R$ decouples from the SM bath, the conventional freeze-out mechanism applies. In this case, the SM bath temperature governs the whole process;
\item Late freeze-out: When dark matter freeze-out occurs after $\nu_R$ decoupling, a more complex scenario emerges. Here, $\nu_R$ and dark matter form a separate thermal bath, necessitating appropriate consideration of both SM and $\nu_R$ bath temperatures in the Boltzmann equations.
\end{enumerate}

The temperature of right-handed neutrinos $T_{\nu_R}$ is expressed as a function of the SM bath temperature ($T_\gamma$), with their decoupling governed by the effective coefficient $G_X$. 
Initially, when temperatures are high, $T_{\nu_R}$ closely tracks $T_\gamma$. As the Universe cools, $\nu_R$ eventually decouples from the SM bath, with the epoch of decoupling determined by $G_X$. A larger $G_X$ value delays the decoupling, resulting in a higher $T_{\nu_R}$ at later epochs relative to scenarios with smaller $G_X$. The Boltzmann equation governing the evolution of comoving DM density can then be expressed as \cite{Gondolo:1990dk}
\begin{equation}\label{eq:BZeqn2}
    \frac{dY_{\chi}}{dx}=-\beta(T_\gamma)\frac{s(T_\gamma)}{\mathcal{H}(T_\gamma,T_{\nu_R},M_{\chi})}\frac{1}{x}\left<\sigma v\right>\left({Y_{\chi}}^2-{(Y_{\chi}^{\rm eq}})^2\right),
\end{equation}
where $x={M_{\chi}}/{T_\gamma}$, $M_{\chi}$ is the mass of $\chi$ and $\left<\sigma v\right>$ is the thermally averaged cross section of $\overline{\chi}\chi\rightarrow\overline{\nu_R}\nu_R$. The other relevant quantities appearing in the above Boltzmann equation can be defined as
{\scriptsize{
\begin{eqnarray}
 Y_\chi = \frac{n_\chi}{s}, \,\,   n_{\chi}(T_{\nu_R},M_{\chi})&=&\frac{g_\chi}{2\pi}T_{\nu_R}M_{\chi}^2K_2\left(\frac{M_{\chi}}{T_{\nu_R}}\right),
\end{eqnarray}
}}
\begin{eqnarray}
 s(T_\gamma)&=&\frac{2\pi^2}{45}g_{*s}(T_\gamma)T_\gamma^3,\\
    \mathcal{H}(T_\gamma,T_{\nu_R},M_{\chi})&=&\sqrt{\frac{8\pi}{3M^2_{\rm pl}}\rho_{\rm Tot}(T_\gamma,T_{\nu_R},M_{\chi})},\\
    \beta(T_\gamma)&=&1+\frac{T_\gamma}{3g_{*s}(T_\gamma)}\frac{dg_{*s}(T_\gamma)}{dT_\gamma},
\end{eqnarray}
\begin{eqnarray}
\sigma(\overline{\chi}\chi\rightarrow\overline{\nu_R}\nu_R)&=&\frac{G'^2_V}{4\pi}\frac{1}{s}\left(1-\frac{4M_{\chi}^2}{s}\right)^{-1/2}\nonumber\\
&{}&\times\left(\frac{s^2}{3}+\frac{2 M_{\chi}^2 s}{3}-2M_{\chi}^4\right). 
\end{eqnarray}
Here $g_\chi$ is the internal degrees of freedom of $\chi$, $g_{*s}(T_\gamma)$ is the effective degrees of freedom, $\rho_{\rm Tot}(=\rho_\gamma(T_\gamma)+\rho_{\nu_R}(T_{\nu_R})+\rho_{\chi}(T_{\nu_R},M_{\chi}))$ is the total energy density, and $M_{\rm pl}$ is the Planck mass ($=1.22\times10^{19}$ GeV). 
\begin{figure}[H]
		\centering
        \includegraphics[scale=0.45]{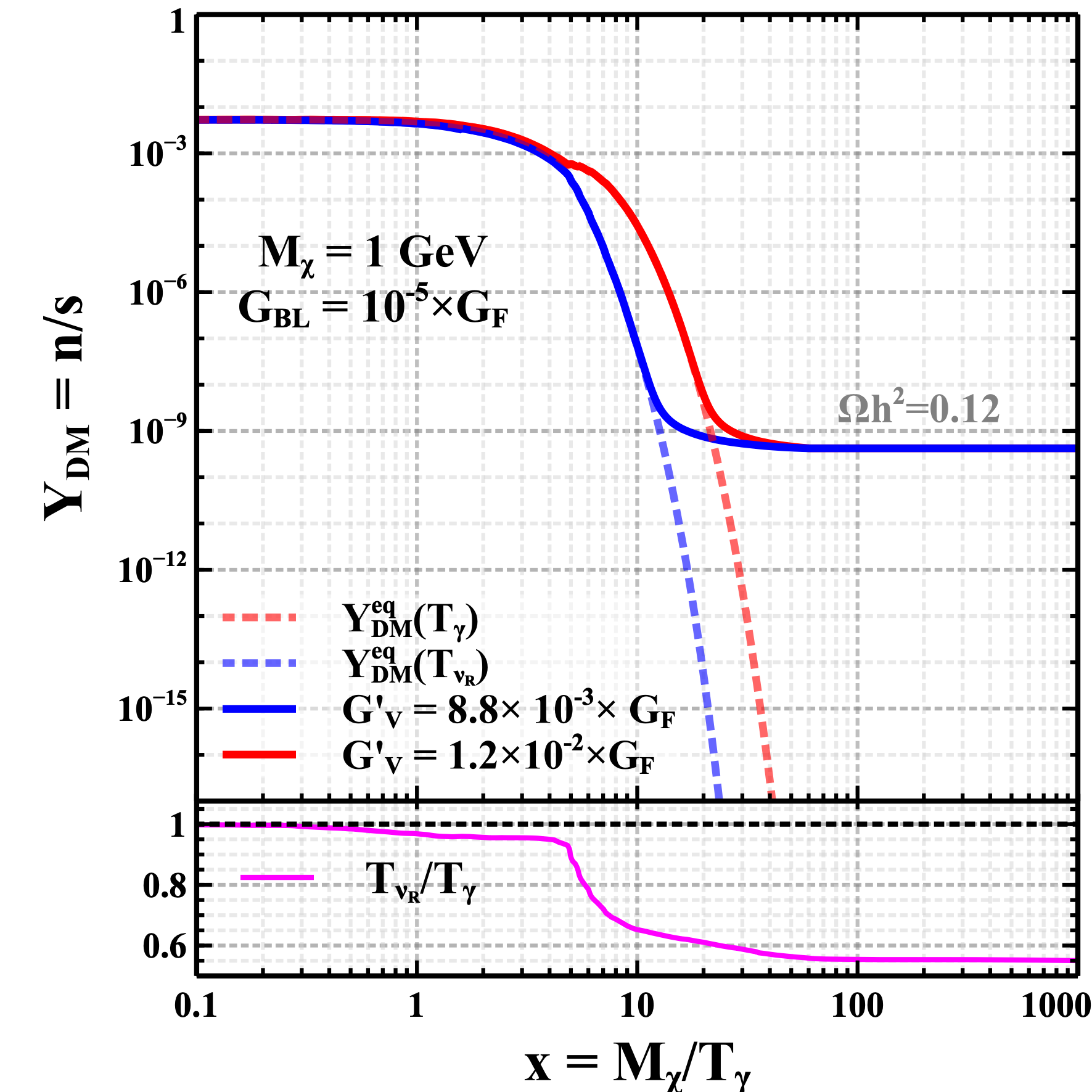}
		\caption{Here we have shown the evolution of DM abundance (top) and $\left(T_{\nu_R}/T_{\gamma}\right)$ (magenta line in the bottom panels) with respect to $x=M_{\chi}/T_{\gamma}$.  The bottom panels depict the $\nu_R$ decoupling from SM thermal bath.} 
		\label{fig:dmabundance}
\end{figure}
Figure \ref{fig:dmabundance} illustrates the evolution of DM number density, obtained by solving the Boltzmann equation [cf. Eq. \eqref{eq:BZeqn2}]. We present results for a benchmark DM mass of 1 GeV. As discussed previously, the DM dynamics are primarily governed by the temperature of the right-handed neutrino ($\nu_R$) bath, which in turn depends on its interaction rate with the SM bath, controlled by the effective operator coefficient $G_X$. The bottom insets of Fig. \ref{fig:dmabundance} depict the corresponding evolution of $T_{\nu_R}$ relative to the photon temperature $T_\gamma$. The magenta line ($T_{\nu_R}/T_\gamma$) deviates from the thick black dotted line ($T_{\nu_R}/T_\gamma=1$), indicating the decoupling of $\nu_R$ from the SM bath.
The blue dashed line represents the equilibrium abundance of DM as a function of $T_{\nu_R}$, while the red dashed line shows the equilibrium DM density as a function of the photon temperature $T_\gamma$. Once $\nu_R$ decouples from the SM bath, these two lines diverge.
We investigated two scenarios for the decoupling of DM. In one scenario, we assumed the $\nu_R$ bath shares the same temperature with SM bath and the red solid line shows the evolution of DM. In contrast, the blue line corresponds to DM decoupling from $\nu_R$ thermal bath where the evolution of $T_{\nu_R}$ is decided by solving Eq. (\ref{eq:tnurevo}). The $G'_V$ is fine-tuned in both cases such that the red and blue lines yield the correct relic density. As shown in the plot, earlier decoupling of the $\nu_R$ bath necessitates earlier decoupling of dark matter from the $\nu_R$ sector, due to Boltzmann suppression of the equilibrium dark matter abundance. Consequently, achieving the correct relic density in this case requires a smaller annihilation cross section.
\iffalse
We investigated two scenarios for the decoupling temperature of $\nu_R$ by solving the Boltzmann equation for DM evolution. The left plot showcases a late decoupling scenario for $\nu_R$, with $G_X (=G_V)=10^{-3}\times G_F$, while the right plot illustrates an earlier decoupling scenario, where $G_X (=G_V)=10^{-5}\times G_F$. In both plots, the green and orange lines correspond to identical $G'_V$ values as mentioned in the inset of the figure. Notably, the earlier $\nu_R$ decoupling scenario (right plot) results in a lower relic density compared to the late $\nu_R$ decoupling scenario (left plot). The purple line in both plots represents the correct relic density, achieved by fine-tuning the $G'_V$ coefficients. As anticipated, the earlier $\nu_R$ decoupling scenario requires a smaller $G'_V$ value to attain the correct relic density, reflecting the need for earlier DM decoupling.
\fi
In Fig. \ref{fig:DM2}, we summarise the correct relic density parameter space in the plane of $\Delta N_{\rm eff}$ and DM mass (lower $x$-axis) ranging from 1 GeV to 1000 GeV, while different colors of solid contours represent the effective coefficient $G_S$ in the {\it left panel} and $G_V$ in the {\it right panel}. For a fixed value of $G_S$ (or $G_V$), the DM mass and $G'_V$ show one-to-one correspondence to provide correct relic density, which is shown in the upper $x$-axis. For simplicity, we consider only vector-type effective interactions for the DM. We constrain the DM mass to be below the EFT scale. The DM annihilation cross section increases with the center-of-mass energy and, consequently, with DM mass~\cite{Cohen:2021gdw}.  As the cross section grows with increasing mass, a smaller effective coefficient is required to achieve the correct relic density. This inverse relationship is clearly demonstrated in the figure, where lower $G'_V$ values correspond to larger DM masses. Similarly, in Fig. \ref{fig:DMT}, we present the DM relic density that satisfies the parameter space for three benchmark values of $G_T$. As shown in Fig. \ref{fig:neffT}, a substantial range of $G_T$ values is ruled out by the NMM bounds. However, there exists a narrow range of $G_T(<3.82\times 10^{-8}G_F)$ where the thermal relic density of $\nu_R$ can contribute minimally to $\Delta N_{\rm eff}$.
\begin{figure*}
		\includegraphics[width=0.43\linewidth]{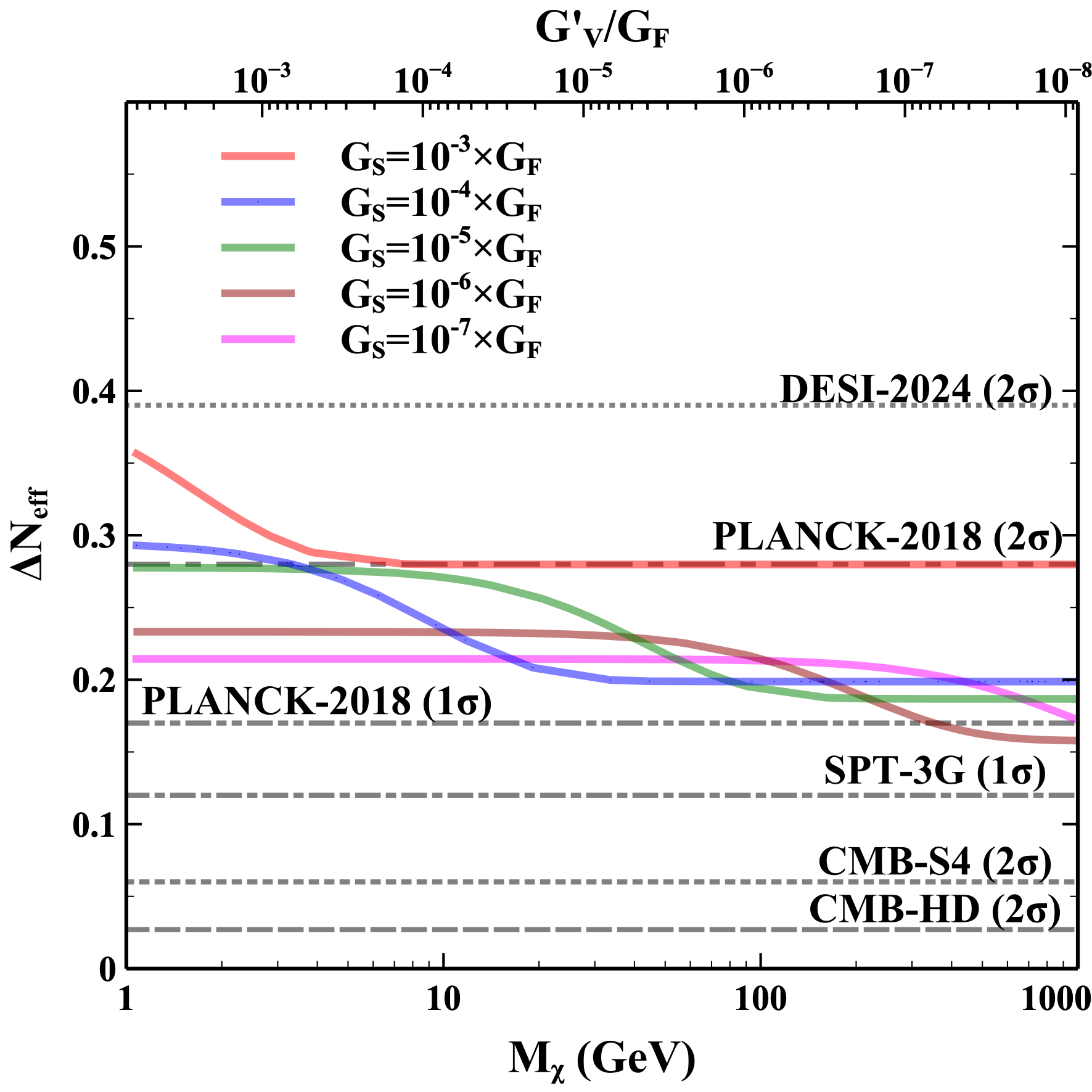}\hfil
		\includegraphics[width=0.43\linewidth]{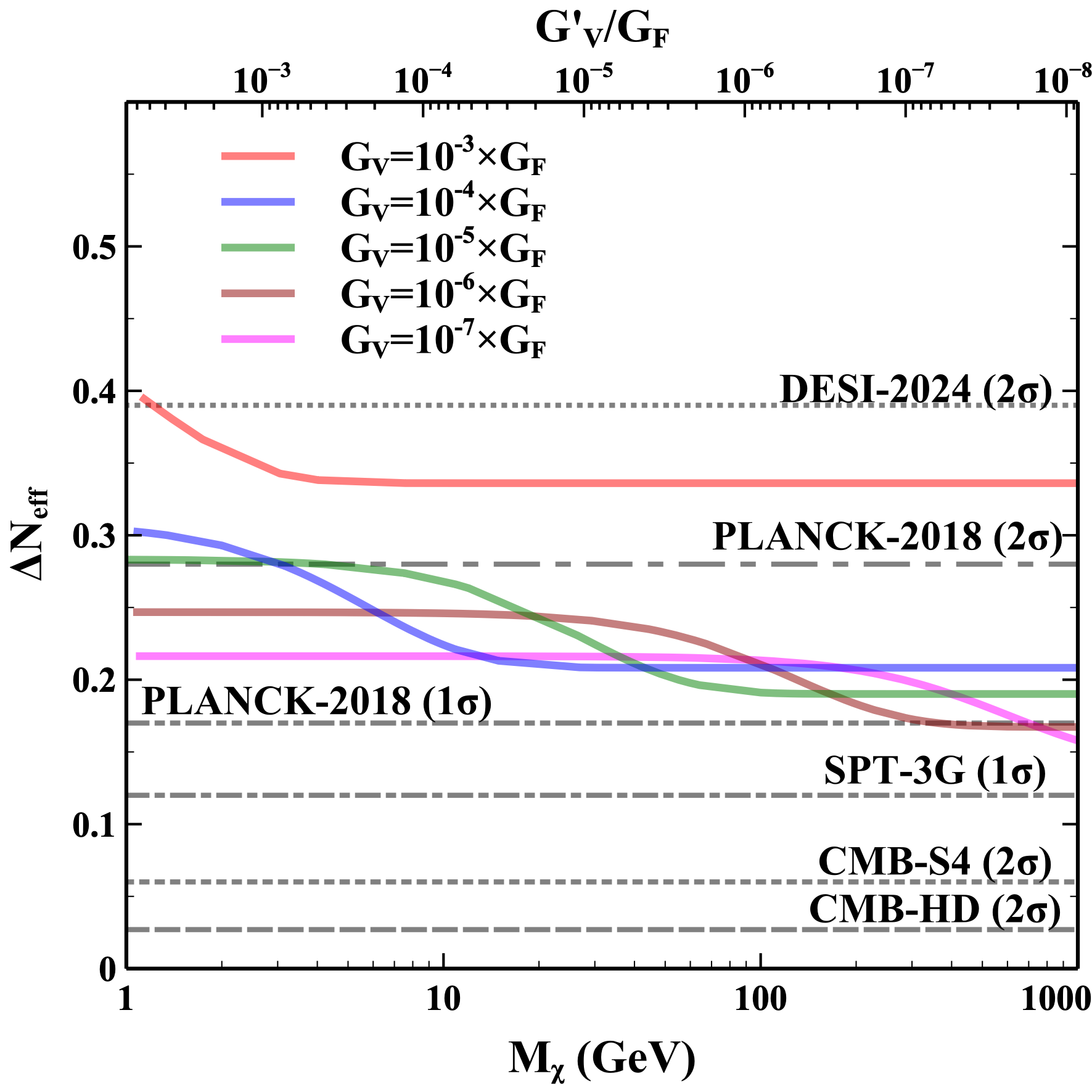}
	\caption{Relic satisfying points are plotted in the plane of DM mass and $\Delta N_{\rm eff}$. In the {\it left} panel, the colored contours represent the corresponding effective coefficient $G_S$, while in the {\it right} panel, they represent the corresponding $G_V$ value.}
	\label{fig:DM2}
\end{figure*}

\begin{figure}[h]
    \centering
   \includegraphics[scale=0.48]{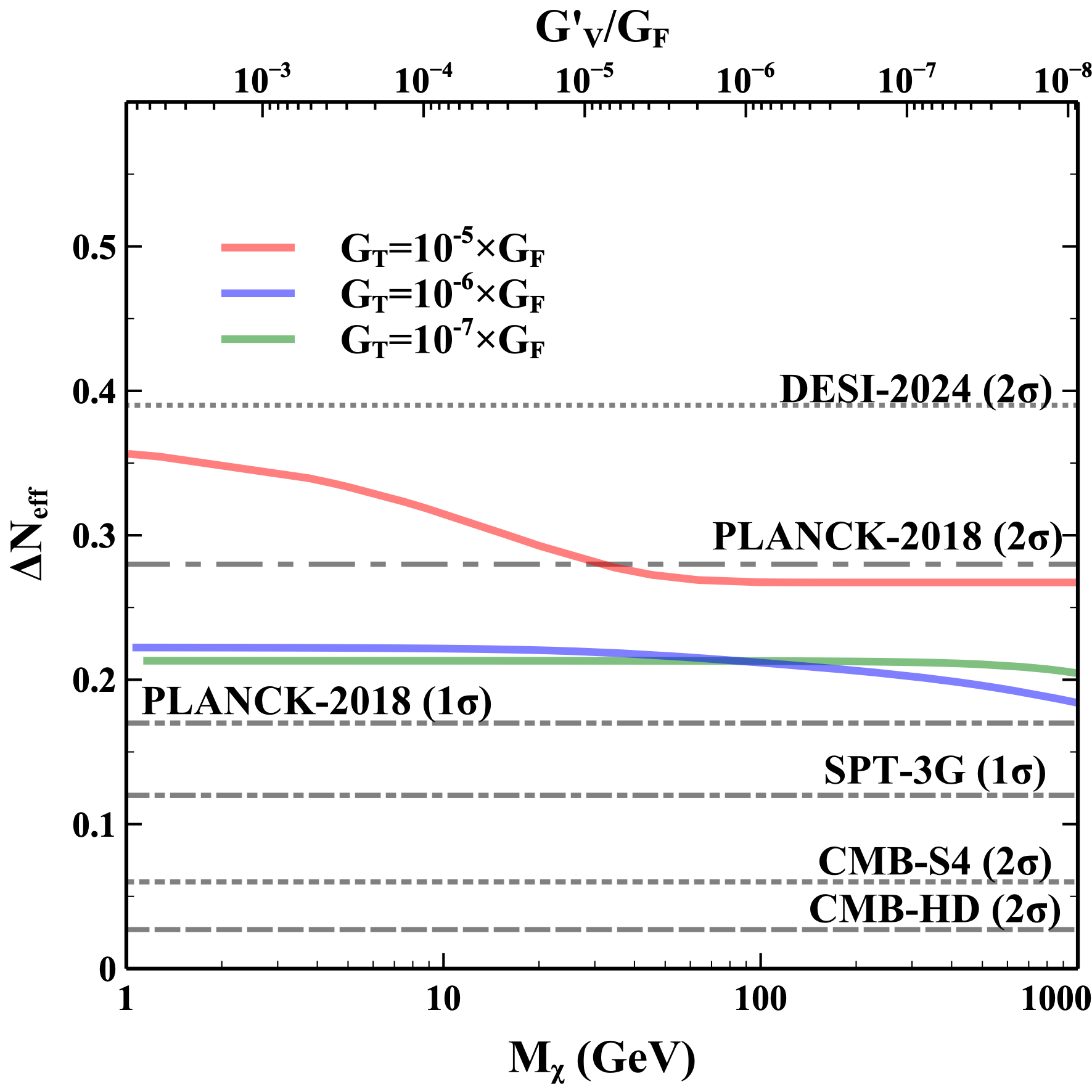}
    \caption{Relic satisfying points are plotted in the plane of DM mass and $\Delta N_{\rm eff}$. The colored contours represent the corresponding $G_T$ value.}
    \label{fig:DMT}
\end{figure}
\vspace{0.5cm}
\item\textit{Direct Detection:}\\
\begin{figure}[h]
		\centering
  \includegraphics[scale=0.1]{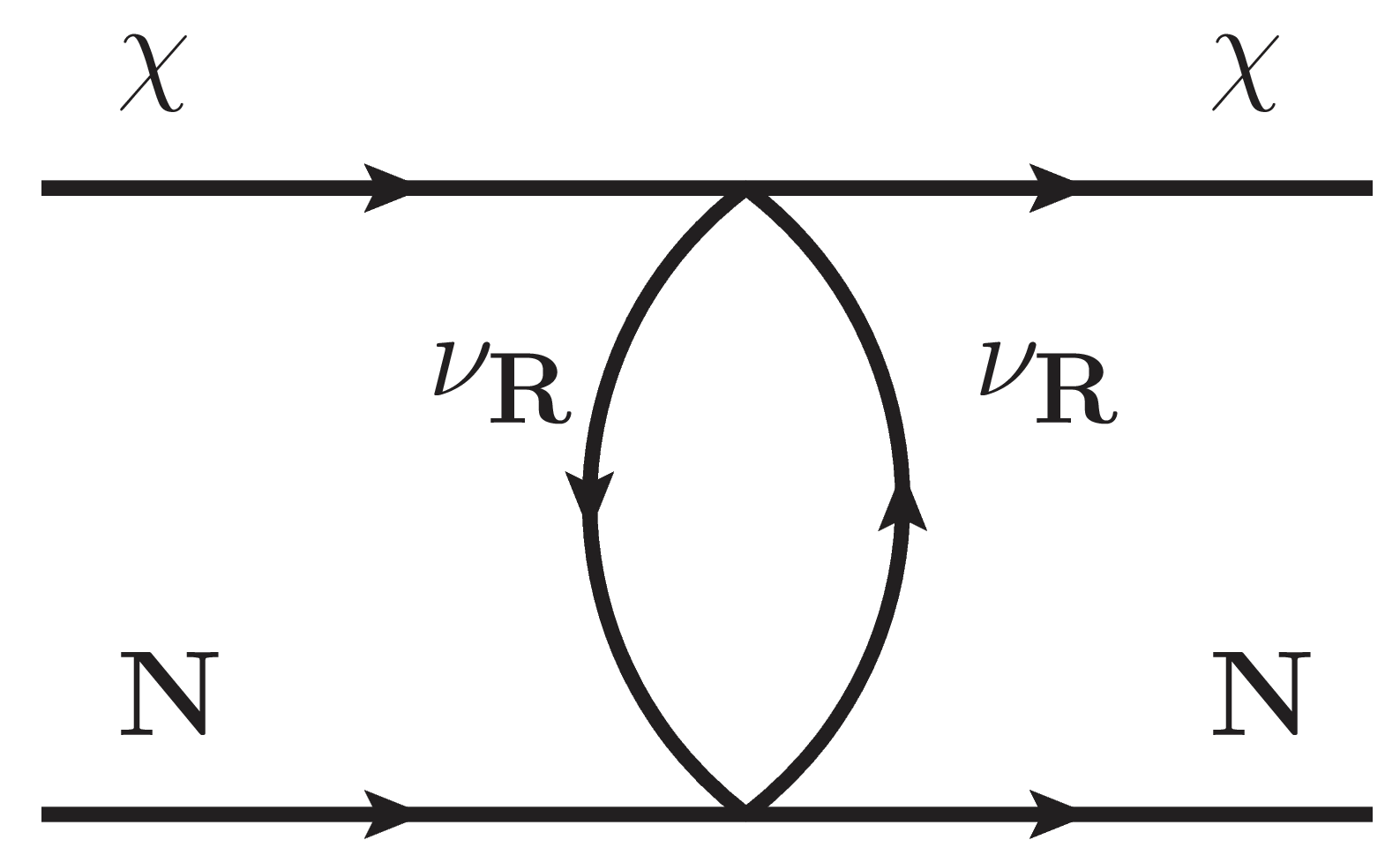}
		\caption{DM-nucleon scattering processes for DM direct detection.}
		\label{fig:c2dddiag}
\end{figure}
\begin{figure}[h]
		\centering
  \includegraphics[scale=0.43]{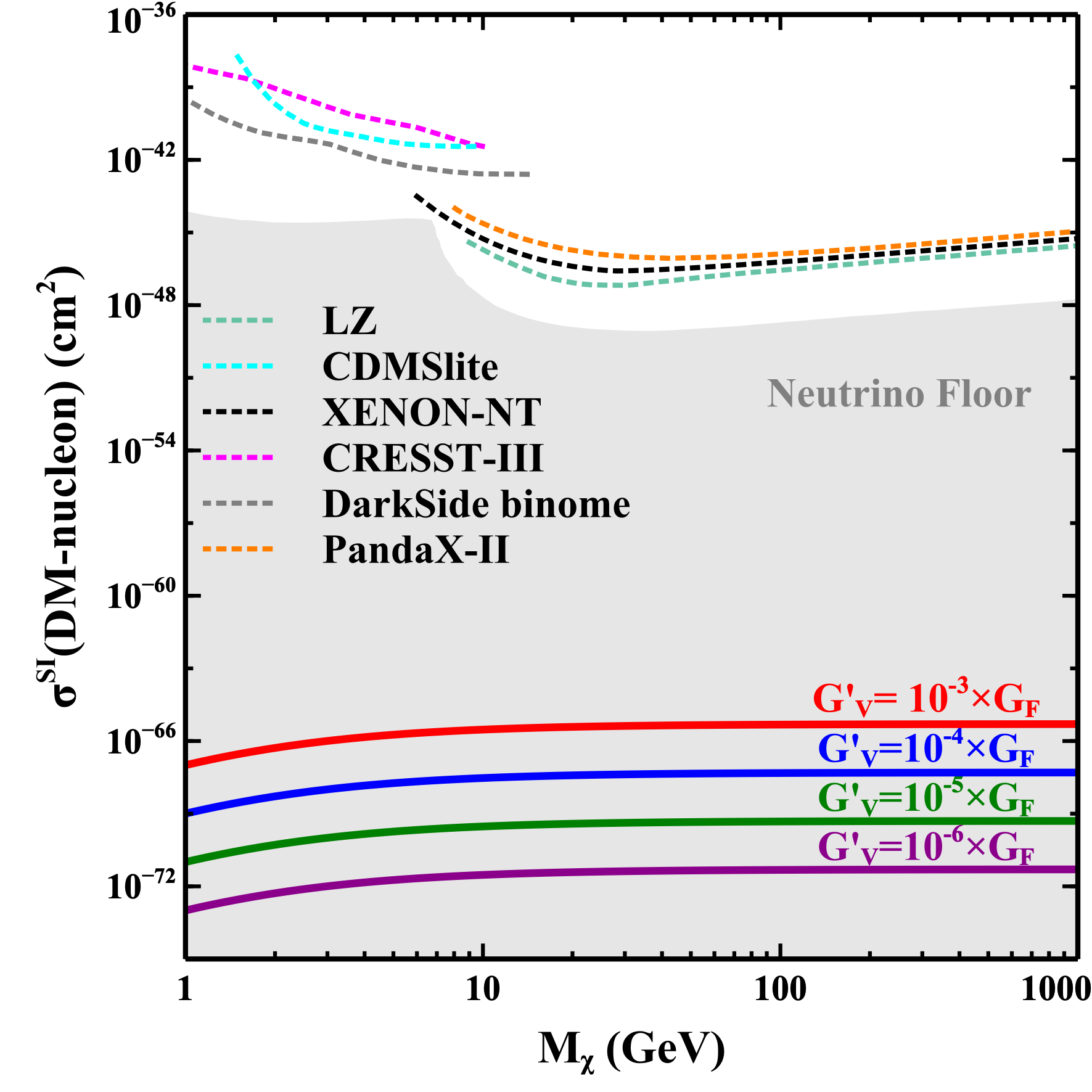}
		\caption{Spin-independent DM-nucleon scattering cross section as a function of DM mass.}
		\label{fig:DM2signature}
\end{figure}
In the $\nu_R$-philic DM model, DM lacks direct interactions with the SM and hence  DM-nucleon scattering is not possible at the tree-level. However, DM direct detection remains feasible through a loop-mediated process involving right-handed neutrinos ($\nu_R$), as illustrated in Fig. \ref{fig:c2dddiag}. The direct detection cross section for loop-level DM-nucleon scattering is given by
\begin{equation}
    \sigma_{\chi N}=\int_0^{4v^2\mu^2_{\chi N}}dp^2\frac{1}{64 \pi}\frac{1}{M^2_{\chi}M^2_{N}}\frac{1}{v^2}\overline{\left|\mathcal{M}_{\chi N}\right|^2},
\end{equation}
where the reduced mass of the DM-nucleon system is given as, $\mu_{\chi N}=\frac{M_{\chi}M_{N}}{M_{\chi}+M_{N}}$. We assume a DM velocity of $v\sim 10^{-3}$ for DM in the galactic halo. The amplitude of the loop diagram $\left|\mathcal{M}_{\chi N}\right|^2$ is detailed in Appendix \ref{app:loopdd}. In Fig. \ref{fig:DM2signature}, we present the spin-independent DM-nucleus scattering cross section as a function of DM mass (thick colored lines) along with the existing experimental constraints (dotted lines). The solid lines represent the DM-nucleus scattering cross section for benchmark values of $G'_V$ scaled with $G_F$, with $G_V$ fixed at $10^{-3} \times G_F$.  The dashed lines show the experimental bounds from XENON \cite{XENON:2023cxc,XENON:2019zpr}, LUX-ZEPLIN(LZ) \cite{LZ:2022lsv}, CDMSlite \cite{SuperCDMS:2018gro}, CRESST-III \cite{CRESST:2019jnq}, DarkSide-50 Binomial Fluctuation \cite{DarkSide:2018bpj}, and PandaX-II \cite{PandaX-II:2017hlx}. The gray shaded region shows the neutrino floor \cite{Billard:2021uyg}.
For each $G'_V$ value, we impose a perturbative-unitarity cutoff on the DM mass to ensure the validity of the EFT treatment. This can be seen from some of the solid contours appearing only below a certain value of DM mass. The loop contribution significantly suppresses the direct detection cross section, placing it far below the present and future sensitivity of DM direct search experiments. Consequently, this $\nu_R$-philic DM scenario could potentially explain the null detection of DM in terrestrial direct detection facilities. \\

\item\textit{Indirect Detection:}\\
\end{enumerate}
In addition to direct detection, our $\nu_R$-philic DM model allows for indirect detection through loop-induced processes involving $\nu_R$. These processes enable DM annihilation to charged fermions, which can be constrained by various gamma-ray and cosmic-ray experiments as well as from CMB anisotropy.
 \begin{figure}[h]
		\centering
  \includegraphics[scale=0.45]{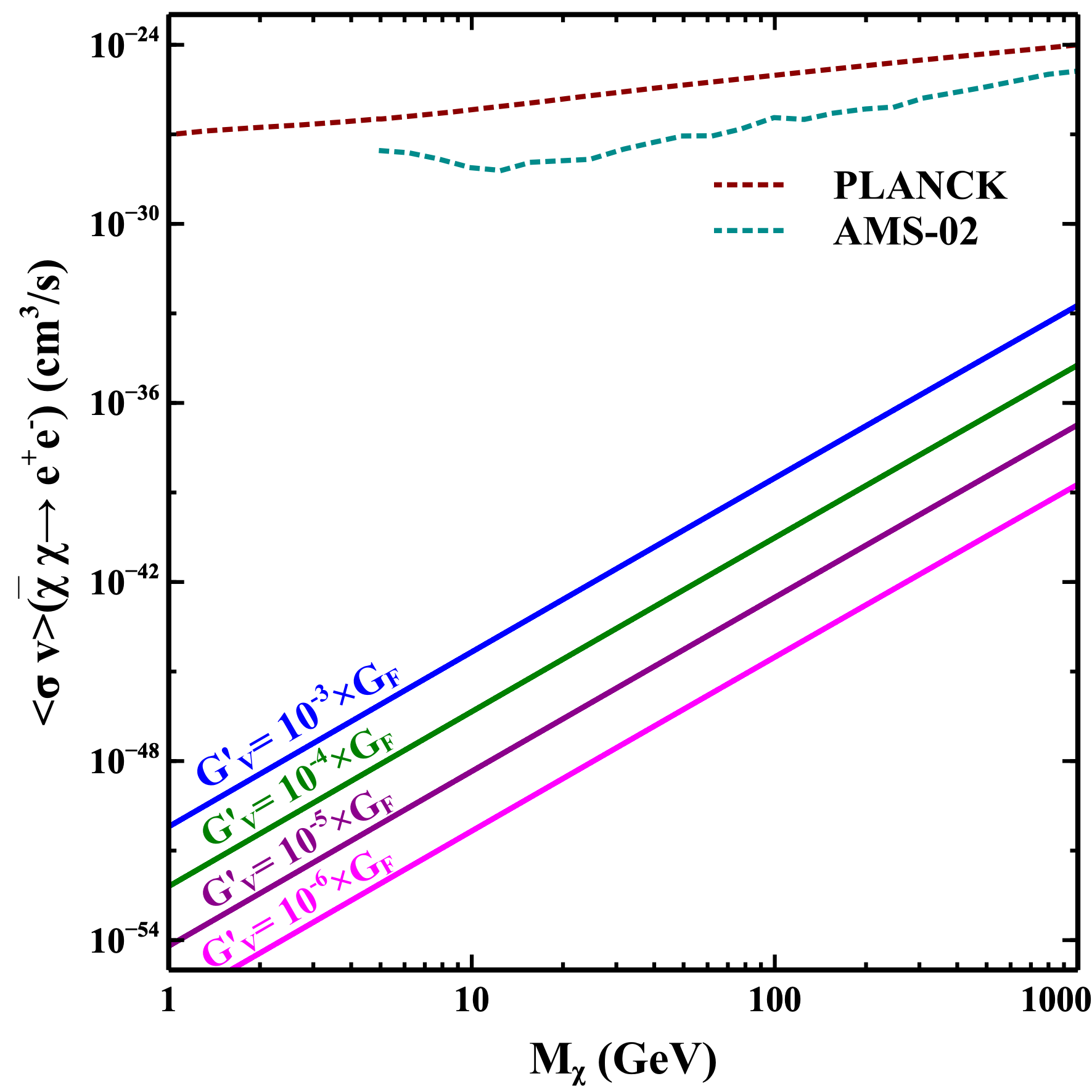}
		\caption{DM annihilation cross section $ \langle\sigma v\rangle_{ \overline{\chi} \chi \to e^+ e^-}$ as a function of DM mass.}
		\label{fig:c2id}
\end{figure}

Figure \ref{fig:c2id} illustrates the thermal averaged annihilation cross section of DM to $e^+e^-$ via the $s$-wave channel as a function of DM mass. We present results for several benchmark values of $G'_V$ scaled with $G_F$. The solid lines represent the annihilation cross section, while the dashed lines show the current observational bounds from CMB measurement \cite{Slatyer:2015jla} and AMS02 \cite{John:2021ugy}. Similar to the case of direct detection, here also we apply the upper bound on DM mass from validity of EFT, as seen for the chosen values of effective couplings in green and blue colored contours.
\section{$U(1)_{B-L}$: An Example of UV Completion}\label{sec:uvmodels}
%\section{UV completion: $U(1)_{B-L}$ Model}\label{sec:uvmodels}
In this section, we extend our effective theory framework of $\nu_R-SM$ and $\nu_R-\chi$ interactions with a UV complete scenario. Gauged $B-L$ extension of the SM \cite{Davidson:1978pm, Mohapatra:1980qe, Marshak:1979fm, Masiero:1982fi, Mohapatra:1982xz, Buchmuller:1991ce} has been a popular BSM framework studied extensively in the literature. Here, we augment SM gauge symmetry with a gauged $U(1)_{B-L}$ symmetry, under which the quarks and leptons carry a charge of $\frac{1}{3}$ and $-1$, respectively. The inclusion of three right-handed neutrinos becomes a natural requirement for anomaly cancellation.
In order to prevent Majorana mass of right-handed neutrinos, one can have an unbroken $U(1)_{B-L}$ with a massive neutral gauge boson, $Z'$, using the Stueckelberg mechanism \cite{FileviezPerez:2019cyn}. Alternatively, one can choose the scalar content in a way which breaks $B-L$ symmetry by more than 2 units to forbid the Majorana mass term \cite{Heeck:2014zfa}. The Dirac mass term can be generated from the usual Yukawa Lagrangian, given by
\begin{equation}
    \mathcal{L}\supset y_{\alpha\beta}\overline{l_L}_\alpha \tilde{H}{\nu_R}_\beta + {\rm H.c.},
\end{equation}
where $\alpha$ and $\beta$ are flavor indices, $l_L$ is the lepton doublet, $\tilde{H}=i\sigma_2H^{*}$. The Yukawa couplings ($y_{\alpha\beta}$) are required to be sufficiently small to generate the observed small neutrino masses, consistent with the neutrino mass scale inferred from experimental data. In addition to the Yukawa interactions, the RHNs couple to SM fermions through gauge interactions, with the exchange of a $Z'$ boson. These interactions govern the decoupling of the $\nu_R$ from the SM bath. Furthermore, we extend the particle content of the SM by introducing a singlet fermion, $\chi$ and a scalar singlet, $\phi$. We propose an additional $Z_2$-symmetry to ensure the stability of the dark sector particles, wherein both $\chi$ and $\phi$ are odd under the $Z_2$ symmetry and all other particles are even. We consider $\chi$ as the lightest $Z_2$ odd particles and thus serves as a fermionic DM candidate. To exhibit the $\nu_R$-philic DM nature, we keep $\chi$ as neutral under SM and $U(1)_{B-L}$ gauge group, whereas $\phi$ is assigned with a $-1$ charge under $U(1)_{B-L}$ symmetry and transforms trivially under SM symmetries. $\nu_R$ acts as a portal between the SM thermal bath and dark sector particles which is dictated by the interaction term in the Lagrangian,
\begin{equation}
    \mathcal{L}\supseteq y_\chi \overline{\nu_R}\chi\phi + H.c.
\end{equation}
 Here it is worth mentioning that, $\phi$ being a $Z_2$ odd particle, does not get vacuum expectation value and ensures the stability of $\chi$. The scalar Lagrangian is given by
\begin{equation}
    \mathcal{L}\supseteq \left(D_\mu \phi\right)^{\dagger}D^\mu \phi - \mu_\phi^2\left(\phi^\dagger \phi\right) - \lambda_\phi \left(\phi^\dagger \phi\right)^2 - \lambda_{\phi H}\left(\phi^\dagger \phi\right)\left(H^\dagger H\right),
\end{equation}
where $D_\mu=\partial_\mu - i g_{BL} Z'_\mu$, $g_{BL}$ is gauge coupling and $Z'$ is the gauge boson for the $U(1)_{B-L}$ gauge symmetry. The inclusion of $\phi$ in the model is merely to have $\chi-\nu_R$ interaction. Therefore, for simplicity, we consider a heavy $\phi$ ($M_\phi=10^4 {\rm~GeV}$), such that the abundance of $\phi$ does not affect the DM phenomenology.

In this model, $\nu_R$ interaccts with SM fermions via exchange of $Z'$ boson and maintains thermal equilibrium with SM bath in the early Universe. As the Universe cools down, the interaction rate drops below the expansion rate resulting a new thermal bath with temperature $T_{\nu_R}(\neq T_\gamma)$. The evolution of $T_{\nu_R}$ can be tracked by solving Eqs. (\ref{eq:evolSM}) and (\ref{eq:evolnuR}), and the corresponding collision term is given as
\begin{equation}
	C^\rho_{\nu_R}=\frac{N_{\nu_R}}{64 \pi^4}\int_{0}^{\infty}ds s^2 \sigma(s)\left[T_\gamma K2\left(\frac{\sqrt{s}}{T_\gamma}\right)-T_{\nu_R} K2\left(\frac{\sqrt{s}}{T_{\nu_R}}\right)\right],
\end{equation}
where
\begin{equation}
	\sigma(s)=\frac{1}{24 \pi}\frac{g_{BL}^4 s}{(M_{Z'}^2-s)^2}.
\end{equation}

\begin{figure}[h]
		\centering
  \includegraphics[scale=0.45]{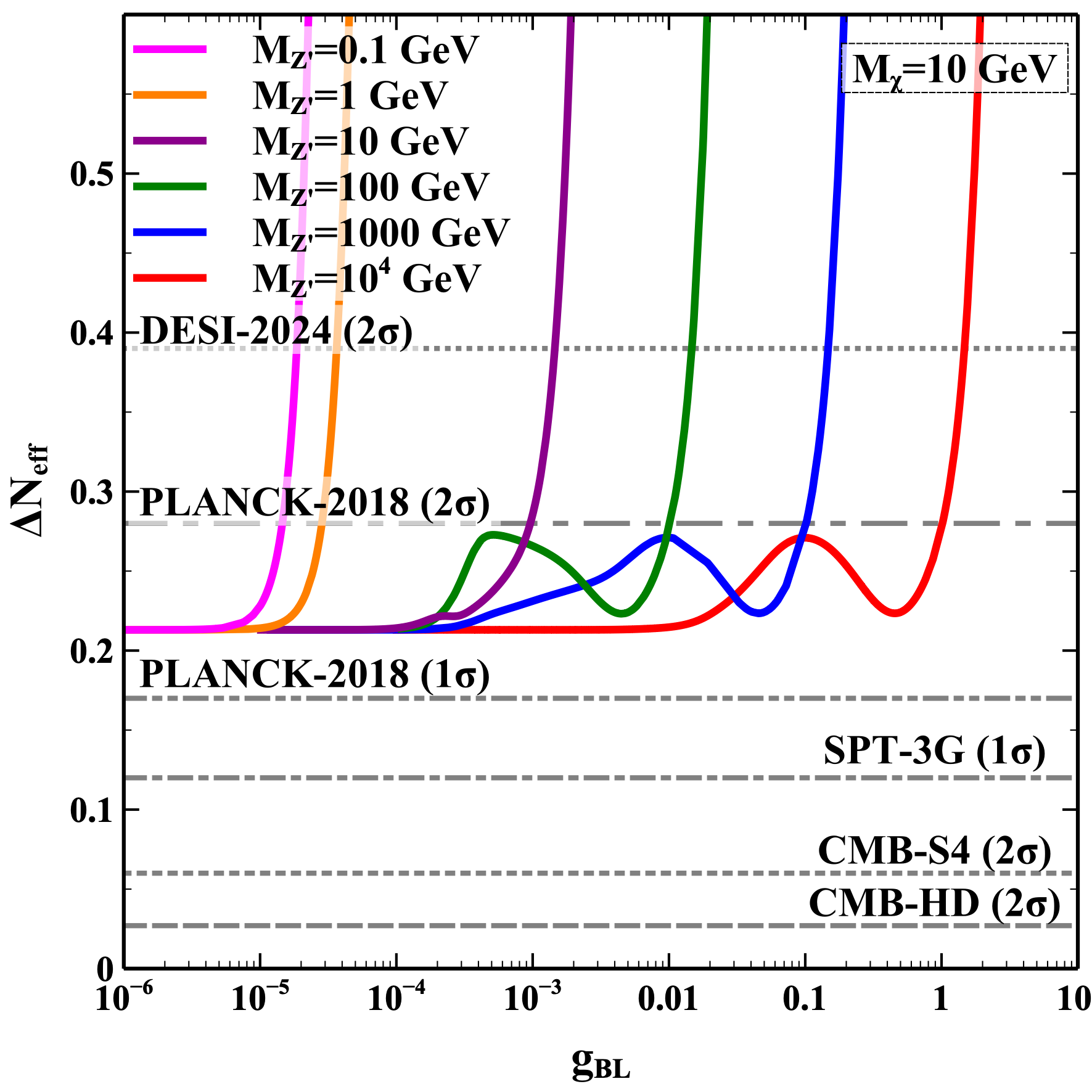}
		\caption{$\Delta N_{\rm eff}$ as a function of $g_{BL}$ for DM mass of 10 GeV. The colored contours correspond to a fixed value of $M_{Z'}$. The current upper bounds and future sensitivities are shown as gray lines.}
		\label{fig:dneff_gbl}
\end{figure}
In Fig.~\ref{fig:dneff_gbl}, we present the variation of $\Delta N_{\rm eff}$ as a function of the gauge coupling $g_{BL}$, with the colored contours corresponding to different values of the $Z'$ boson mass, $M_{Z'}$. For each contour, there exists a minimum thermal contribution to $\Delta N_{\rm eff}$ up to a certain value of $g_{BL}$. Beyond this point, increasing $g_{BL}$ results in a delayed decoupling of the right-handed neutrino thermal bath, leading to a higher value of $\Delta N_{\rm eff}$. Similarly, a lower $M_{Z'}$ enhances the interaction rate, thus postponing the decoupling of the $\nu_R$ bath. This trend is evident in the plot, where the contours corresponding to smaller $M_{Z'}$ exhibit an earlier rise in $\Delta N_{\rm eff}$ along the $g_{BL}$ axis.
\begin{figure}[h]
		\centering
  \includegraphics[scale=0.45]{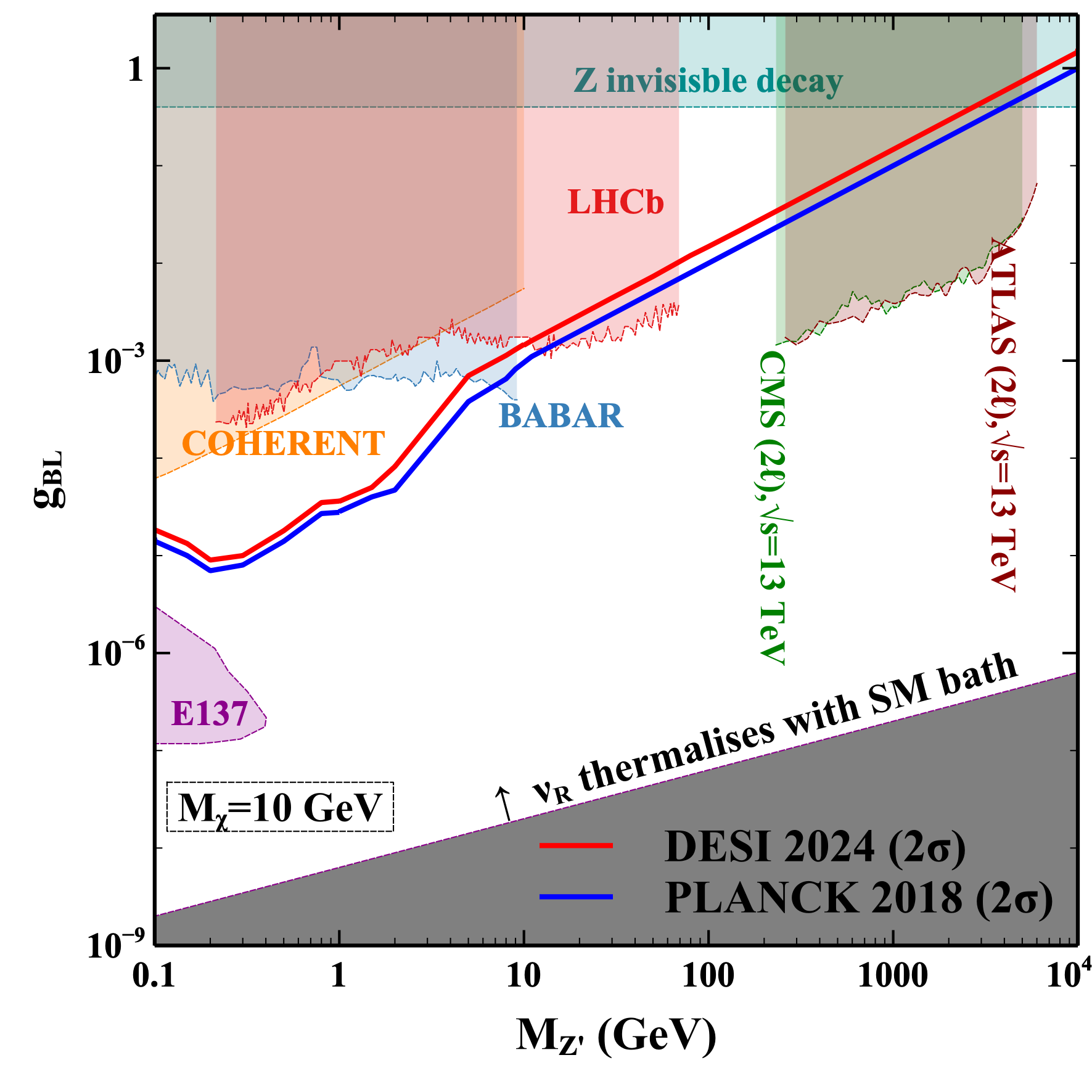}
		\caption{The allowed parameter space in the $g_{BL}~ \text{vs}~M_{Z'}$ plane. The red (blue) solid line represents the constraint on $\Delta N_{\rm eff}$ from the DESI 2024 (PLANCK 2018) results at the $2\sigma$ CL. The gray shaded region represents the out of equilibrium condition for $\nu_R/Z'$.}
		\label{fig:bl_space}
\end{figure}

\begin{figure}[h]
	\centering
	\includegraphics[scale=0.45]{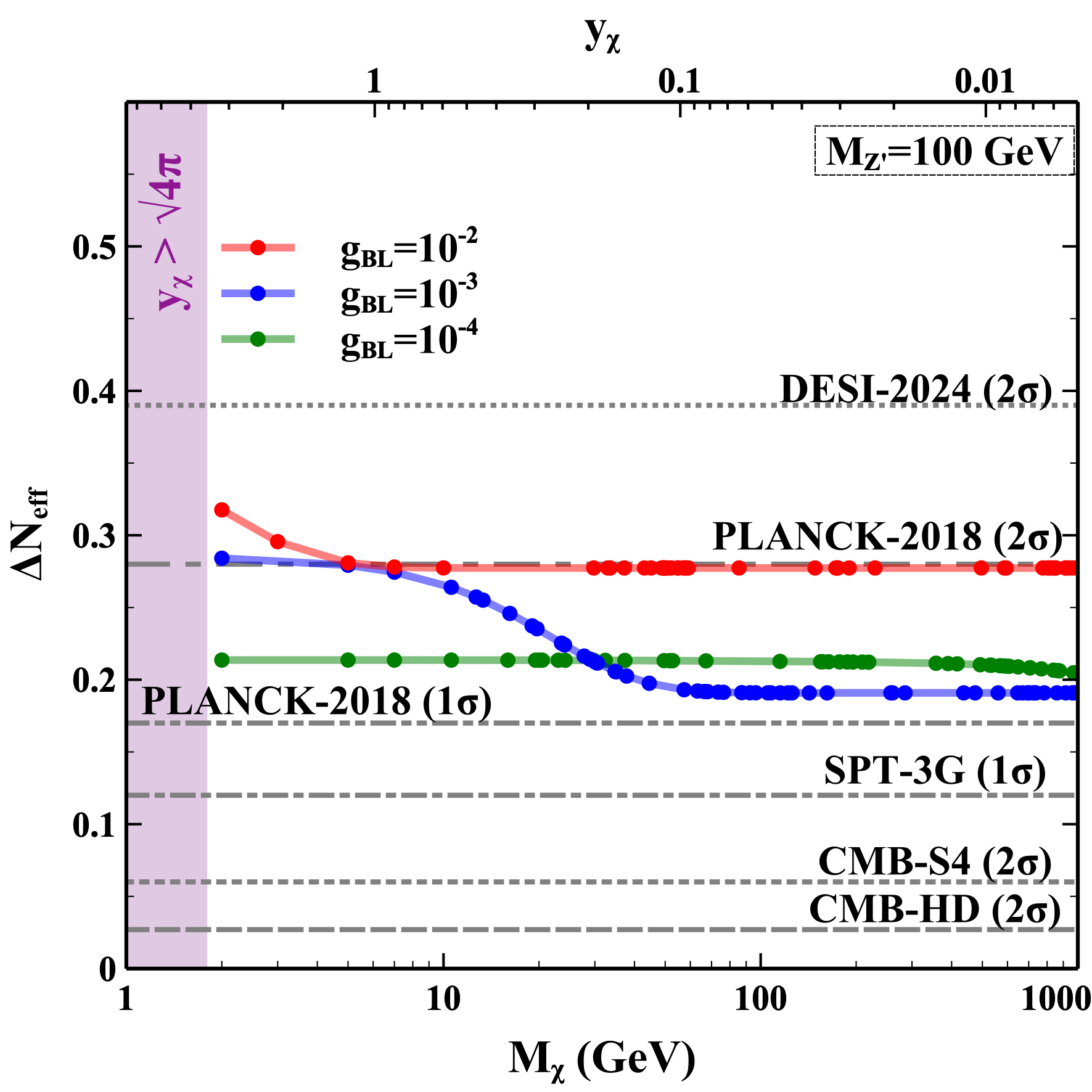}
	\caption{$\Delta N_{\rm eff}$ as a function of DM mass $M_\chi$ consistent with correct relic abundance and fixed $Z'$ mass. The upper $x$-axis indicates the corresponding Yukawa coupling, $y_\chi$. The colored contours represent different values of the gauge coupling, $g_{BL}$. The purple shaded region indicates the perturbative bound on the Yukawa coupling $y_\chi$.}
	\label{fig:DM_relic}
\end{figure}
In Fig. \ref{fig:bl_space}, we present the allowed parameter space in $M_{Z^\prime} - g_{BL}$ plane based on the constraints coming from DESI 2024 \cite{DESI:2024mwx,Allali:2024cji} and PLANCK 2018 \cite{Planck:2018vyg}, indicated by the red and blue colored solid lines, respectively. The shaded regions correspond to the updated constraints coming from various other experiments, such as CMS \cite{CMS:2021ctt}, ATLAS \cite{ATLAS:2019erb}, LHCb\cite{LHCb:2019vmc}, BABAR \cite{BaBar:2014zli}, COHERENT \cite{Cadeddu:2020nbr}, E137 \cite{Bjorken:1988as} and $Z$ invisible decay \footnote{In this model, $Z$ can mix with $Z'$ boson naturally via SM fermions in the loop. By adding all the one-loop contributions, we have estimated the natural mixing parameter, $\epsilon=0.08 g_{\rm BL}$. Consequently, $Z$ can decay to $\nu_R$. Using the constraint from invisible decay of $Z$ boson, we have shown the upper bound on $g_{\rm BL}$ in cyan colored shade in Fig. \ref{fig:bl_space}.} \cite{ParticleDataGroup:2024cfk}, on the gauge coupling ($g_{BL}$) for the $Z'$ boson mass, within the range of $0.1$ GeV to $10^4$ GeV.  The region above gray shaded region represents the $g_{BL}-M_{Z'}$ value for which the $\nu_R$ can come to equilibrium with SM bath.

In this model, the DM relic is established by DM annihilation to $\nu_R$ mediated by $\phi$ particle via $t$-channel diagram. The corresponding cross section is provided in Appendix \ref{app:dmannihilation}. Figure~\ref{fig:DM_relic} illustrates the parameter space yielding the correct DM relic abundance, where the colored contours correspond to different values of the gauge coupling $g_{BL}$, as indicated in the legend. In this analysis, we fix the $Z'$ boson mass at 100 GeV. The vertical axis displays the resulting values of $\Delta N_{\rm eff}$, while the lower horizontal axis represents the DM mass. The corresponding Yukawa coupling $y_\chi$, required to achieve the observed relic density, is shown on the upper horizontal axis. It is worth noting that the behavior observed here closely mimics the trend exhibited in the left panel of Fig.~\ref{fig:DM2}, derived within the EFT framework.
\begin{figure}[h]
	\centering
	\includegraphics[scale=0.45]{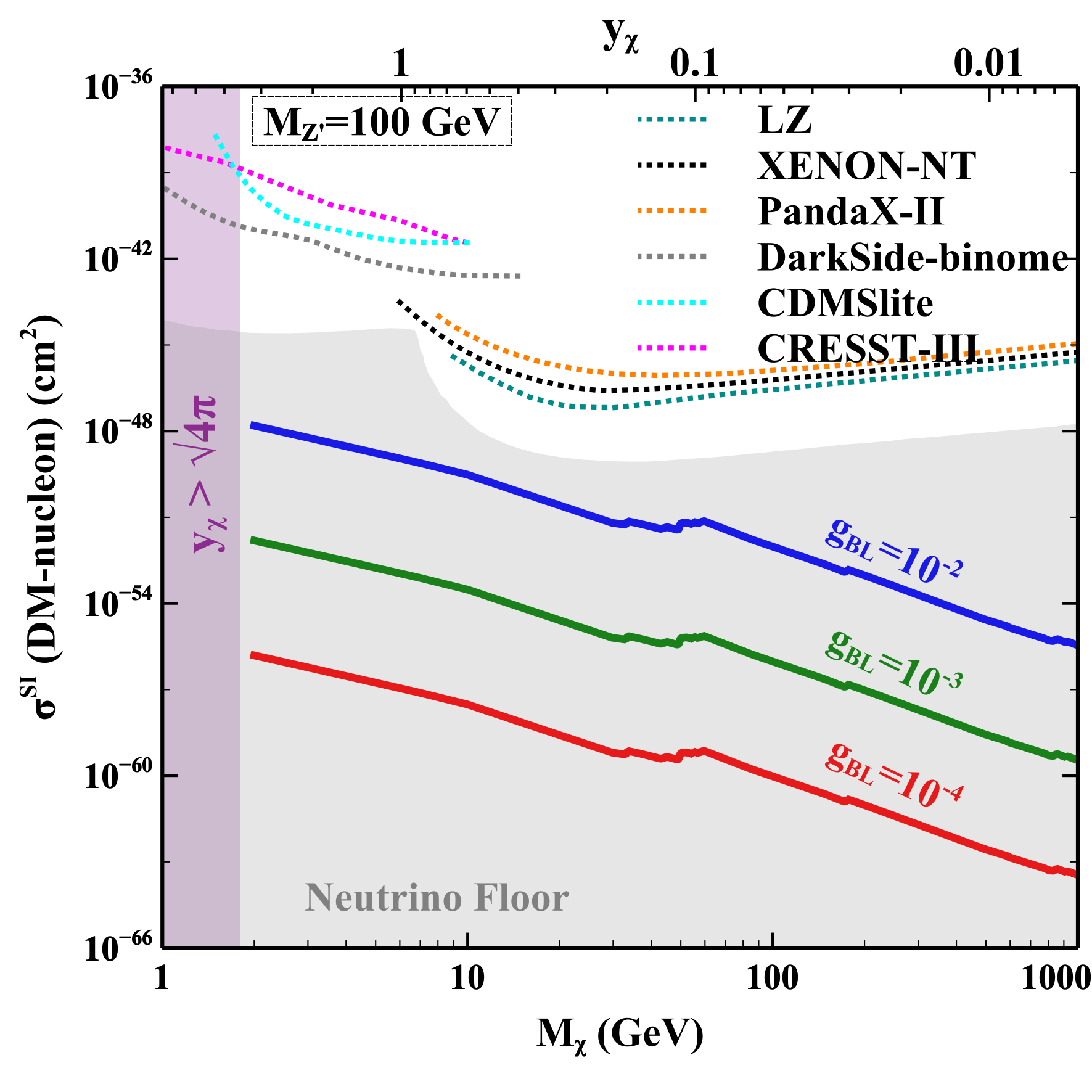}
	\caption{Direct detection cross section corresponding to the parameter space that yields the correct relic density, as a function of the DM mass. The associated Yukawa couplings, $y_\chi$, are indicated on the upper $x$-axis. The colored contours denote different values of the gauge coupling $g_{BL}$. Existing experimental bounds from direct detection searches are presented as dotted colored lines, while the gray shaded region indicates the neutrino floor.}
	\label{fig:Direct_ditection}
\end{figure}
In this model, DM does not interact with Standard Model particles at tree level. However, it can contribute to direct detection signals via two distinct loop-mediated processes. One of which is mediated by Higgs and the other diagram is mediated by $Z'$ boson. Here, the Higgs-mediated diagram remains suppressed because of small coupling, so we have considered only the $Z'$-mediated diagram in this analysis. In Fig.~\ref{fig:Direct_ditection}, we display the direct detection cross section for parameter points consistent with the observed relic abundance, considering DM masses in the range of 1 GeV to 1000 GeV. The dotted lines represent current experimental limits from direct detection searches. 

Since our DM sector couples only to $\nu_R$, direct detection via nuclear or electron recoils lies entirely below the neutrino floor and is not expected to yield observable signals in current or near-future weakly interacting massive particle searches. However, this feature could provide a plausible explanation for the absence of DM signals in terrestrial detectors. Notably, cosmological probes such as measurements of the effective number of relativistic degrees of freedom $N_{\rm eff}$, and effects on small-scale structure caused by DM–$\nu$ scattering offer complementary tests of $\nu_R$-portal DM models. Precision CMB/BAO data and Lyman‑$\alpha$ constraints can thus play a critical role in probing this framework which is otherwise invisible to direct detection experiments. Since our model predicts $\Delta N_{\rm eff} \geq 0.21$, any future nonobservation of such dark radiation can falsify our setup.

\section{Conclusions}\label{sec:conclusion}
In this work, we have studied an effective field theory of light Dirac neutrinos and dark matter, assuming the latter to interact with the standard model only via right chiral parts of Dirac neutrinos. Assuming DM to be a vectorlike singlet fermion stabilized by an unbroken $Z_2$ symmetry, we systematically write down effective operators of the lowest possible dimension involving DM-$\nu_R$ and $\nu_R$-SM interactions. This naturally connects the thermalization of DM with that of $\nu_R$ leading to enhancement of the effective relativistic degrees of freedom $N_{\rm eff}$, within reach of future CMB experiments. We found that the relic abundance of dark matter is intricately tied to the temperature evolution of the $\nu_R$ bath, which is further influenced by its interactions with the SM bath. This allows one to distinguish two distinct freeze-out scenarios, each with unique implications for the dark matter parameter space. We also find enhancement of $N_{\rm eff}$ in the presence of both DM-$\nu_R$ and $\nu_R$-SM interactions compared to the scenarios without DM discussed in earlier works. While typical direct and indirect detection prospects of such DM remain suppressed in agreement with null results, future CMB experiments bring complementary detection avenues.

Finally, we studied UV complete $U(1)_{B-L}$ gauge extension with light Dirac neutrinos. This model naturally incorporate right-handed neutrinos thereby accommodating light Dirac neutrinos and offer interesting complementarity between detection prospects at CMB and collider experiments. Inclusion of DM in such setups can further enhance $\Delta N_{\rm eff}$ improving the detection prospects.\\\\

\textbf{ACKNOWLEDGMENTS}\\\\
The work of D.B. is supported by the Science and Engineering Research Board (SERB), Government of India Grants No. MTR/2022/000575 and No. CRG/2022/000603. D.B. also acknowledges the support from the Fulbright-Nehru Academic and Professional Excellence Award 2024-25. S.M. acknowledges the financial support from National Research Foundation (NRF)
grant funded by the Korea government (MEST) Grant No. NRF-2022R1A2C1005050. The work of D.N. is partially supported by JSPS Grant-in-Aid for JSPS Research Fellows No. 24KF0238. S.K.S. would like to thank Xun-Jie Xu for valuable discussions and insights.
\\\\
\textbf{DATA AVAILABILITY}\\\\
The data tha support the findings of this article are not publicly available. The data are avaialable from the authors upon reasonable request.

\section*{}
\appendix\label{appendix}
\section{POSSIBLE $\nu_R$ INTERACTIONS WITH SM BATH AND THEIR COLIISION TERMS}\label{app:EFTnuR}

Following the four-fermion effective terms in Eq. (\ref{eq:effLag2}), the energy transfer between the $\nu_R$ and SM baths happen through the interactions,
\begin{eqnarray}\label{eq:nuRSMprocess}
    \nu_{R} + \overline{\nu_R} &\leftrightarrow& f_L + \overline{f_L},\label{eq:nuRSMprocess2}\\
    \nu_R + f_L &\leftrightarrow& \nu_R + f_L,\label{eq:nuRSMprocess3}\\
    \nu_R + \overline{f_L} &\leftrightarrow& \nu_R + \overline{f_L},\label{eq:nuRSMprocess4}\\
   \nu_R + \overline{\nu_L} &\leftrightarrow& \overline{f}+ f.\label{eq:nuRSMprocess5}
\end{eqnarray}
The collision terms for the scalar and vector type interactions are given in Table \ref{tab:collision}, and for tensorial interactions, we have followed the prescription given in \cite{Biswas:2024gtr}.
\begin{table*}[t]
    \centering
    \begin{tabular}{||cc|c|cc||}
    \hline
    \hline
 & process & $C_{\nu_{R}}^{(\rho)}$ from MB statistics & $1-\delta_{{\rm FD}}$ & \tabularnewline
\hline 
 & $\nu_{R}(p_{1})+\overline{\nu_{R}}(p_{2})\leftrightarrow\nu_{L}(p_{3})+\overline{\nu_{L}}(p_{4})$ & $\frac{2}{\pi^{5}}| {G}_{S}-2G_{V}|^{2}N_{\nu_{R}}\left(T_{{\rm SM}}^{9}-T_{\nu_{R}}^{9}\right)$ & 0.8841 & \tabularnewline
 & $\nu_{R}(p_{1})+\nu_{L}(p_{2})\leftrightarrow\nu_{R}(p_{3})+\nu_{L}(p_{4})$ & $\frac{1}{2\pi^{5}}| {G}_{S}-2G_{V}|^{2}N_{\nu_{R}}T_{{\rm SM}}^{4}T_{\nu_{R}}^{4}\left(T_{{\rm SM}}-T_{\nu_{R}}\right)$ & 0.8518 & \tabularnewline
 & $\nu_{R}(p_{1})+\overline{\nu_{L}}(p_{2})\leftrightarrow\nu_{R}(p_{3})+\overline{\nu_{L}}(p_{4})$ & $\frac{3}{\pi^{5}}| {G}_{S}-2G_{V}|^{2}N_{\nu_{R}}T_{{\rm SM}}^{4}T_{\nu_{R}}^{4}\left(T_{{\rm SM}}-T_{\nu_{R}}\right)$ & 0.8249 & \tabularnewline
 \hline
\end{tabular}
    \caption{Collision terms $C_{\nu_{R}}^{(\rho)}$.}
    \label{tab:collision}
\end{table*}

\section{LOOP CONTRIBUTION IN DIRECT DETECTION}\label{app:loopdd}
The loop factor, for the diagram given in Fig. \ref{fig:c2dddiag}, is given by
\begin{equation}
    I=\frac{G_VG'_V}{(2\pi)^4}\int d^4k \frac{\cancel{k}}{k^2+i\epsilon}\frac{\cancel{p}+\cancel{k}}{(p+k)^2+i\epsilon}.
\end{equation}
The integration above diverges for large loop momentum, so we employ dimensional regularization to regularize the integration.
In $n$-dimensions, the integration takes the form
\begin{eqnarray}
    I&=&\frac{G_VG'_V\mu^{4-n}}{(2\pi)^n}\int d^nk \frac{\cancel{k}}{k^2+i\epsilon}\frac{\cancel{p}+\cancel{k}}{(p+k)^2+i\epsilon}\\
    &=&\frac{nG_VG'_V\mu^{4-n}}{(2\pi)^n}\int d^nk \frac{p.k+k^2}{(k^2+i\epsilon)((p+k)^2+i\epsilon)},\label{eq:loop2}
\end{eqnarray}
where $k$ is the loop momentum, $p$ is the momentum transfer from $\chi$ to nucleus and $\mu$ is chosen to have mass dimension 1. Using the Feynman parameter, $x$, and replacing $k\rightarrow k-px$, one can write the Eq. (\ref{eq:loop2}) as
\begin{equation}
    I=\frac{nG_VG'_V\mu^{4-n}}{(2\pi)^n}\int^1_0 dx \int d^nk \frac{k^2-\overline{Q^2}}{(k^2+\overline{Q^2})^2},
\end{equation}
where $\overline{Q^2}=x(1-x)p^2$. After applying Wick's rotation ($d^nk\rightarrow id^nl_E=idl_El^{n-1}_Ed\Omega_{n-1}=id\Omega_{n-1}\frac{1}{2}dl^2_El^{n-2}_E,~k^2\rightarrow -l_E^2~{\rm and}~p^2\rightarrow -p_E^2$) and simplifying
\begin{equation}\label{eq:loopwick}
    I=\frac{-inG_VG'_V\mu^{4-n}}{{(2\pi)}^n}\frac{2\pi^{n/2}}{\Gamma{(n/2)}}\frac{1}{2}\int^1_0 dx \int_0^\infty dl^2_E \frac{l_E^{n-2}(l_E^2-Q^2)}{(l_E^2+Q^2)^2},
\end{equation}
where $Q^2=x(1-x)p_E^2$ and we used $\int d\Omega_{n-1}=\frac{2\pi^{n/2}}{\Gamma{n/2}}$. For further simplification, we used variable replacement $\frac{l^2_E}{Q^2}\rightarrow\xi$ and the Eq. (\ref{eq:loopwick}) can be read as
\begin{equation}\label{eq:loopbeta}
    I= \frac{-inG_VG'_V\mu^{4-n}}{{(2\pi)}^n}\frac{\pi^{n/2}}{\Gamma{(n/2)}}\int^1_0 dx Q^{n-2}\int_0^\infty d\xi\frac{\xi^{\frac{n}{2}-1}(\xi-1)}{(\xi+1)^2}
\end{equation}
\begin{equation}\label{eq:loopbeta2}
    ~~=iG_VG'_V\left[\frac{p^2_E}{8\pi^2\epsilon}+\frac{p^2_E}{8\pi^2}\left(\gamma_E-\frac{3}{2}+\ln\frac{p^2_E}{4\pi\mu^2}\right)+\mathcal{O}(\epsilon)\right],
\end{equation}
where $\gamma_E$ is defined as Euler's constant and carries a value of 0.577216 and $n=4-2\epsilon$. While solving from Eq. (\ref{eq:loopbeta}) to Eq. (\ref{eq:loopbeta2}), we used the definition of $\beta$-function, $\beta(m+1,n+1)=\int^\infty_0 \frac{du u^m}{(1+u)^{m+n+2}}$. We note that the loop factor can be regularized for any finite value of $\epsilon$. Now, applying Wick's rotation on Eq. (\ref{eq:loopbeta2}),
\begin{equation}\label{eq:effcoupling}
    I=-iG_VG'_V\left[\frac{p^2}{8\pi^2}\left(\gamma_E-\frac{3}{2}+\ln\left|\frac{p^2}{4\pi\mu^2}\right|-i\pi\right)+\mathcal{O}(\epsilon)\right]
\end{equation}
Equation (\ref{eq:effcoupling}) represents the effective coefficient of the DM-nucleus scattering through a loop mediated process. For practical purposes, one may choose the $\mu$ value such that the logarithmic term can be neglected.
\begin{equation}
    \overline{\left|\mathcal{M}_{\chi N}\right|^2} = 16\left(Z\alpha_p+(A-Z)\alpha_n\right)^2 M^2_{\rm DM}M^2_{N}\left|I\right|^2
\end{equation}
where $A,Z$ are mass number and atomic number, respectively, $\alpha_{p,n}\simeq 3$ and $M_n$ is the nucleon mass.

\section{DM ANNIHILATION CROSS SECTION}\label{app:dmannihilation}
\begin{widetext}
{\scriptsize{
\begin{eqnarray}
    \sigma(\overline{\chi}\chi\rightarrow\overline{\nu_R}\nu_R)=&&-\frac{y_{\chi }^4}{32 \pi  s \left(s-4 M_{\chi }^2\right)}\times
    \left(4 \left(M_{\chi }-M_{\phi }\right) \left(M_{\chi }+M_{\phi }\right) \log \left(\frac{s \left(\sqrt{\frac{s-4 M_{\chi }^2}{s}}-1\right)+2 M_{\chi }^2-2 M_{\phi }^2}{-s \left(\sqrt{\frac{s-4 M_{\chi }^2}{s}}+1\right)+2 M_{\chi }^2-2 M_{\phi }^2}\right)\right.\nonumber\\
    &&~~~~~~~~~~~~~~~~~~~~~~~~~~~~~~~~~~~~~~~~~~~~~~~~~~~~~~~\left. -\frac{2 s \sqrt{\frac{s-4 M_{\chi }^2}{s}} \left(s M_{\phi }^2+2 \left(M_{\chi }-M_{\phi }\right) \left(M_{\chi }+M_{\phi }\right) \left(M_{\chi }^2-M_{\phi }^2\right)\right)}{s M_{\phi }^2+M_{\chi }^4-2 M_{\chi }^2 M_{\phi }^2+M_{\phi }^4}\right)
\end{eqnarray}
}}

\end{widetext}
\twocolumngrid
\bibliographystyle{utphys}
%\bibstyle{apsrev}
%\bibliography{ref}	
\providecommand{\href}[2]{#2}\begingroup\raggedright

\endgroup

\end{document}